\documentclass[final,5p,times,twocolumn,numbers,sort,compress]{elsarticle}
\usepackage{amssymb}
\usepackage{makeidx}
\usepackage{amsfonts}
\usepackage{amsmath}
\usepackage{eurosym}
\usepackage{graphicx}
\usepackage{tensor}
\usepackage[utf8]{inputenc}
\usepackage{color}
\usepackage{revsymb}
\usepackage{url}

\usepackage{hyperref}
 \hypersetup{
     colorlinks=true,
     linkcolor=blue,
     filecolor=blue,
     citecolor = blue,
     urlcolor=cyan,
     breaklinks=true
     }
 \usepackage[capitalize]{cleveref}

\def\be{\begin{equation}}
\def\ee{\end{equation}}
\def\bea{\begin{eqnarray}}
\def\eea{\end{eqnarray}}

\def\l{\left}

\begin{document}

\begin{frontmatter}

\title{Testing Weyl geometric gravity with the SPARC galactic rotation
curves database}

\author[inst1]{Maria Cr\u{a}ciun}
\affiliation[inst1]{organization={`Tiberiu Popoviciu' Institute of Numerical Analysis, Romanian Academy},
            addressline={57, Fantanele Street}, 
            city={Cluj-Napoca},
            postcode={400320},
            state={Cluj},
            country={Romania},
            }

\author[inst2,inst3,inst4]{Tiberiu Harko\corref{cor1}}
\cortext[cor1]{Corresponding author. \ead{tiberiu.harko@aira.astro.ro}}

\affiliation[inst2]{organization={Department of Theoretical Physics, National Institute of Physics
and Nuclear Engineering (IFIN-HH)},
            addressline={},
            city={Bucharest},
            postcode={077125},
            state={},
            country={Romania}}
\affiliation[inst3]{organization={Department of Physics, Babes-Bolyai University},
            addressline={Kogalniceanu Street},
            city={Cluj-Napoca},
            postcode={400084},
            state={Cluj},
            country={Romania}}
\affiliation[inst4]{organization={Astronomical Observatory, Romanian Academy},
            addressline={19 Ciresilor Street},
            city={Cluj-Napoca},
            postcode={400487},
            state={Cluj},
            country={Romania}}

\begin{abstract}
We present a detailed investigation of the properties of the galactic rotation curves in the
Weyl geometric gravity model, in which the gravitational action is
constructed from the square of the Weyl curvature scalar, and of the strength
of the Weyl vector. The theory admits a scalar-vector tensor representation,
obtained by introducing an auxiliary scalar field. By assuming that the Weyl
vector has only a radial component, an exact solution of the field equations
can be obtained, which depends on three integration constants, and, as compared to the Schwarzschild solution, contains two new terms, linear and quadratic in the radial coordinate. In the framework of this solution we obtain the exact general relativistic expression
of the tangential velocity of the massive test particles moving in stable circular
orbits in the galactic halo. We test the theoretical predictions of the model by using 175 galaxies from the
Spitzer Photometry \& Accurate Rotation Curves (SPARC) database. We fit the
theoretical predictions of the rotation curves in conformal gravity with the
SPARC data by using the Multi Start and Global Search methods. In the total expression of the tangential velocity we also include the effects of the baryonic matter, and  the mass to luminosity ratio. Our results indicate that the simple solution of the Weyl geometric gravity can successfully
account for the large variety of the rotation curves of the SPARC sample, and provide a satisfactory description of the particle dynamics in the galactic halos, without the need of introducing the elusive dark matter particle.
\end{abstract}


\end{frontmatter}

\tableofcontents

\section{Introduction}

The recent results of the analysis of the Cosmic Microwave Background
radiation \cite{Pl1} have confirmed the existence of a good agreement
between the observational data and the standard, spatially-flat 6-parameter $%
\Lambda$CDM cosmological model, having a power-law spectrum of adiabatic
scalar perturbations. A combined analysis of the polarization, temperature,
and lensing data gives a dark matter density $\Omega_ch^2 = 0.120 \pm 0.001$%
, and a baryonic matter density $\Omega _bh^2 = 0.0224 \pm 0.0001$,
respectively. Moreover, the basic inferred (model-dependent) present day
value of the Hubble function is given by $H_0 = (67.4 \pm 0.5)$ \;km/ s/
Mpc, while the matter density parameter and the matter fluctuation
amplitudes are obtained as $\Omega_m = 0.315 \pm 0.007$ and $\sigma _8 =
0.811 \pm 0.006$ \cite{Pl1}, respectively. Hence, baryonic matter
constitutes only around 17\% of the total mass budget in the Universe, the
rest of the matter being in the form of dark matter.

Dark matter resides in
large halos around the visible baryonic matter distribution in galaxies, and
its existence is inferred from two important observational evidences: the
behavior of the galactic rotation curves, and the mass deficit in clusters
of galaxies.

Despite its remarkable success in explaining cosmological observations, the $%
\Lambda$CDM model is facing several important challenges when trying to
explain the astrophysical properties of the galaxies, and of their dark matter halos. The discrepancies between
theory and observations become more important for the dwarf galaxies, with
the $\Lambda$CDM theoretical predictions for the number, spatial
distribution, and internal structure of low-mass dark-matter halos being
contradicted by the observed properties of dwarf galaxies \cite{Sales}. One
of these problems is the core-cusp problem, which originated in the
numerical simulations suggesting dark matter densities diverging as $\rho
\propto r^{-1}$ in the inner regions of the halo \cite{NFW}. On the other
hand, observations of the rotation curves in some dwarf galaxies indicate
that their inner densities are consistent with a constant-density core \cite%
{Oh}. This problem can be solved by assuming, for example, that dark matter
is in the form of a Bose-Einstein Condensate \citep{BEC1,BEC2,BEC3}.

A persistent, and a yet unsolved challenge to the $\Lambda$CDM model is the observed
diversity of the rotation curves \cite{Sales}. The observed rotation curves
of different types of galaxies show a wide range of forms. Even the rotation
curves have a similar outer behavior, their inner behavior is still very
different. Simulations of baryonic matter could not consistently describe
the different velocity curve shapes in the inner regions of the galaxies
\cite{Sales}.

Up to now, the only evidence for the existence of dark matter is only
gravitational, and no experimental detection of any dark matter particle has
been reported yet. Hence, the problem of the reality of the dark matter particle  is still
open, and the existence of alternative explanations for the observational
data cannot be excluded a priori. One of the possible solutions for the understanding of the
galactic dynamics may be obtained by assuming that dark matter is just a
modification of the gravitational force at galactic or extra-galactic
scales. Hence, beyond the boundaries of the Solar System, the Newtonian or
general relativistic laws of gravity are naturally modified, and the
gravitational phenomena are described by a new, fundamental theory of gravity.

The behavior of the rotation curves may be thus explained by modifying the laws of Newtonian
physics, as done is the MOND (Modified Newtonian Dynamics) theory \cite{r1}.
Modified theories of gravity have been extensively applied to look for
possible explanations of the dark matter related phenomenology \cite%
{r2,r3,r4,r5,r6,r7,r8,r8a,r9,r10,r11,r12,r13,r14,r15,r16}. For example, in \cite{r6} it was
shown that a slight modification of the Einstein-Hilbert Lagrangian of the
form $R^{1+\delta}$, $\delta << 1$, could explain the behavior of the
galactic rotation curves without postulating the existence of dark matter. A
detailed review of the dark matter problem in modified theories of gravity
with geometry-matter coupling is presented in \cite{r17}. For a review of
the particle physics aspects of dark matter see \cite{r18}.

In 1918, a few years after the proposal of general relativity, Weyl \cite%
{Weyl1,Weyl2} introduced a generalization of Riemann geometry, based on the
idea of the conformal invariance of the physical laws. Weyl's main goal was to
obtain a unified theory of gravitation and electromagnetism. Even as a
unified field theory Weyl's approach is no longer seen as valid, the
geometrical (and physical) ideas he introduced represent an attractive
theoretical framework, on which extensions of general relativity can be
constructed. For detailed presentation of the role of Weyl geometry in
physics see \cite{Weyl3}. Gravitational theories derived from Weyl geometry
are pure metric, and they contain the equivalence principle, as well as the
general covariance principle of general relativity. Moreover, a
supplementary symmetry, local conformal invariance, is added in a nontrivial
way to the theory, by requiring that the action is invariant with respect to
local conformal transformations of the metric given by $g_{\mu\nu}(x)%
\rightarrow e^{2\omega(x)}g_{\mu\nu}(x)$, where the local phase $\omega(x)$
is an arbitrary function of the coordinates. Moreover, Weyl's geometry is non-metric, and has the basic property that the
covariant divergence of the metric is nonzero. This property also leads to the existence of a specific Weyl connection, which generalizes the Levi-Civita connection of Riemannian geometry. For a discussion between the differences between the notions of conformal and Weyl invariance see \cite{Kar}.

One of the modified theories of gravity in which the galactic rotation
curves have been extensively investigated is the conformal Weyl gravity \cite%
{M1,M2,M3,M4,M5,M6}. This modified gravity theory is built on the principle
of local conformal invariance, which severely restricts the choice of the
action for the gravitational field, by requiring that the action remains
invariant under any conformal transformation $g_{\mu \nu} \rightarrow \Omega
(x)g_{\mu \nu}$ of the metric. One of the simplest possibilities to satisfy
the principle of the covariant invariance is to construct the action from
the conformally invariant Weyl tensor $C_{\mu \nu \rho \kappa}$ \cite{M1}.
Once the conformal invariance is strictly imposed, particle masses can only
be generated through the spontaneous breaking of the symmetry of the action.
The complete, exact exterior solution for a static, spherically symmetric
object in conformal  Weyl gravity was found in \cite{M1}. The solution contains as a
particular case the Schwarzschild solution of general relativity, and also
contains in the metric a new term that grows linearly with distance. It was
suggested in \cite{M1} that this solution could provide an explanation for
the observed galactic rotation curves, without the requiring the existence
of dark matter. This interesting idea was further investigated through a
detailed comparison of the theoretical solution with the observations \cite%
{M7,M8,M9,M10, M11}.

The contributions of the Weyl gravity approach to the
galactic dynamics can be summarized as follows. There are two new effects
that do appear on galactic and extra-galactic scales, leading to a
modification of the Newtonian gravity. Locally, the baryonic matter sources
within galaxies generate not only Newtonian potentials, but also linear
potentials. Globally, two new potentials, one linear, and the other one
quadratic, are created by the rest of the ordinary matter in the Universe
\cite{M7}. The universal linear potential term has the form $V(r)=\gamma
_0c^2r/2$, where $\gamma _0$ is a constant, and it can be associated with
the cosmic background itself. The second, de Sitter type universal
potential, is taken as $V(r)=-\kappa c^2 r^2/2$, and it is induced by the
inhomogeneities in the Cosmic Microwave Background Radiation \cite{M7}. In
\cite{M8} Weyl gravity theory was applied to a sample of 111 spiral
galaxies, consisting of high surface brightness galaxies, low surface
brightness galaxies, and dwarf galaxies, respectively, having rotation curve
data points extending beyond the optical disk. By considering as free
parameters only the galactic mass-to-light ratios, the theory can describe
the properties of this set of rotation curves without the need for invoking
the presence of dark matter. The investigations initiated in \cite{M8} were
extended in \cite{M9} by considering a supplementary set of 27 galaxies, of
which 25 are dwarf galaxies, plus 3 additional galaxies belonging to the
original sample. Fully acceptable fits were found for this sample, thus
bringing to 138 the number of rotation curves of galaxies that could be
explained in the conformal gravity theory. These studies seem also to
confirm the idea that dark matter is just a universal contribution to
galactic dynamics, originating from matter located outside of the galaxies,
and thus independent of them.

On the other hand, in \cite{Hob} it was shown that in conformally invariant
gravity theories, defined on Riemannian spacetimes, and having the
Schwarzschild - de Sitter metric as a solution of the Einstein field
equations, the trajectories followed by baryonic matter particles are the
timelike geodesics of the Schwarzschild-de Sitter metric, thus leading to
rotation curves with no flat regions. Moreover, attempts to model rising
rotation curves by fitting the coefficient of the quadratic term for each
independent galaxy cannot be successful, since this term can be interpreted
as a (very small) cosmological constant $\Lambda$. Moreover, it was shown
that the invariance of particle dynamics with respect to the choice of the
conformal frame is also valid for arbitrary metrics. The same results apply
for conformally invariant gravity theories constructed in more general
Riemann-Weyl-Cartan spacetimes. The above results can be
illustrated as follows. In the case of a static spherically symmetric metric
with $g_{00}=1/g_{11}=1-2GM/R-kr^2$, the tangential velocity of massive
particles can be obtained as $v^2=\left(GM/r-kr^2\right)/\left(1-2GM/r-kr^2%
\right)$. Since in the Newtonian limit the two terms in the numerator
dominate, we obtain $v^2=v_{Kepl}^2-kr^2$, where $v_{Kepl}^2=GM/r$. By
assuming $k=\Lambda/3$, with $\Lambda=10^{-52}\;\mathrm{m}^{-2}$, for a
galaxy of mass $M\sim 10^{11}M_{\odot}$, the rotation curve will fall for
all $r$ until the circular orbits become unbounded at a radius $r =
\left(3GM/\Lambda c^2\right)^{1/3} \sim 0.5$ Mpc \cite{Hob}. Hence, it turns
out that the rotation curves do not have a flat region.

An interesting perspective on Weyl geometry, and its physical applications was recently introduced in \cite{Gh1a,Gh2a,Gh3a,Gh4a,Gh5a,Gh6a,Gh7a,Gh8a,Gh9a,Gh10a}. The basic, and novel idea, is to linearize in the curvature scalar the conformally invariant quadratic Weyl action by introducing an auxiliary scalar field. Hence, the initial purely vector-tensor gravity is transformed into a scalar-vector-tensor theories, which has many attractive features. In its linear representation in Weyl quadratic gravity a spontaneous breaking of the D(1) symmetry takes place, triggered by a Stueckelberg type
type, mechanism. As a result, the Weyl gauge field acquires mass from the spin-zero mode of the $\tilde{R}^2$
term in the action. Through this mechanism, from the Weyl action  the Einstein-Proca Lagrangian is reobtained, after the elimination of the auxiliary scalar field.
The Planck scale is generated from the scalar field mode, and, moreover, the cosmological constant also emerges in the broken phase. It also turns out that all the mass scales,
like the Planck scale, the cosmological constant, and the Higgs field originate from geometry \cite{Gh8a}, with the Higgs field generated by the fusion of Weyl bosons
in the early Universe

Various astrophysical and cosmological implications of the scalar-vector-tensor Weyl theory, also called Weyl geometric gravity, have been investigated in \cite{I1, I2,I3,I4,I5,I6}. Black hole solutions in Weyl geometric gravity have been studied numerically in \cite{I3}, where an exact solution of the field equations has also been obtained, corresponding to a specific choice, and form, of the Weyl vector. This solution is mathematically similar to the exact solution of the conformal Weyl gravity \cite{M1}, but its physical meaning, origin and interpretation are different. The possibility that this solution, extended to galactic scales, may account for the description of the behavior of the galactic rotation curves was suggested in \cite{I4}, where a small sample of seven galaxies was used to fit the Weyl geometric theoretical model with the observations.

It is the goal of the present paper to continue, and extend the investigation initiated in \cite{I4} by considering a full comparison of the simple three-parameter Weyl geometric dark matter model with the rotation curves data of the SPARC dataset. The SPARC (Spitzer Photometry \& Accurate Rotation Curves) database \cite{S1,S2} consists of a sample of 175 nearby galaxies, with surface photometry at 3.6 $\mu$m. It also contains high-quality rotation curves, obtained from  HI/H$\alpha$ data. SPARC covers a large  range of galactic morphologies, ranging  from S0 to Irr, galactic luminosities ( 5 dex ), as well as surface brightnesses at 4
dex. Galactic mass models based on SPARC data have also been constructed, and with their help the
ratio (Vbar/Vobs) of baryonic-to-observed velocity  has been quantitatively estimated for different characteristic galactic radii, and various values at [3.6] of the stellar mass-to-light ratio (M/L) \cite{S1,S2}. The SPARC database was extensively used to analyze different dark matter models, and to obtain constraints on the model parameters \cite{S3,S4,S5,S6,S7,S8,S9,S10,S11, S12}.

In order to perform the comparison between the model and the observations we obtain first the full general relativistic expression of the tangential velocity of massive test particles moving in stable circular orbits. In total velocity of the particles we include, together with the Weyl contribution, the effects of the different components of the baryonic matter, together with their corresponding mass to light ratios. Our results indicate that the Weyl geometric gravity dark matter model could offer a satisfactory explanation of galactic dynamics without invoking the need of the existence of dark matter.

The present paper is organized as follows. We briefly review the basic concepts and ideas of Weyl geometry and of Weyl geometric gravity in Section~\ref{sect1}. We present the derivation of the exact solution of the static, spherically symmetric field equations of Weyl geometric gravity in Section~\ref{sect2}, where also the expression of the tangential velocity in the considered metric is given, and its properties are discussed. The results of the fitting of the SPARC data with the theoretical model are presented in Section~\ref{sect3}, where we also discuss the correlations between the parameters of the model and different astrophysical quantities. Finally, we discuss and conclude our results in Section~\ref{sect4}.

\section{Essentials of Weyl geometry, and of Weyl geometric gravity}\label{sect1}

In the present Section we very succinctly present the fundamentals of the
Weyl geometry. Then, we introduce the basic, conformally invariant Weyl action, and we present its linearization in the curvature scalar with the use of an auxiliary scalar field, thus transforming the initial vector-tensor theory into a scalar-vector tensor theory.  The full set of field equations of the Weyl geometric gravity theory, obtained by varying the action with respect to the metric, is also written down.

\subsection{Weyl geometry}

Weyl geometry is based on two fundamental ideas. The first important property of Weyl geometry is that the length of a vector is allowed to vary during parallel transport. Hence, the length $l $ of an arbitrary vector parallelly transported from the point $x^{\mu }$ to the infinitesimally  closed point
$x^{\mu }+\delta x^{\mu }$, the length of a vector changes according to the rule
\begin{equation}
\delta l =l \omega _{\mu }\delta x^{\mu },
\end{equation}%
where by $\omega _{\mu }$ we have denoted the Weyl vector field. Moreover, Weyl geometry has a second important
property, namely,  the extension of the metricity
condition $\nabla _{\alpha }g_{\mu \nu }=0$ of the Riemannian geometry. In Weyl geometry one introduces a new fundamental geometric quantity, called
the nonmetricity $Q_{\lambda \mu \nu }$ and defined with the help of the covariant
derivative of the metric tensor, given by
\begin{equation}  \label{nmetr}
\tilde{\nabla}_{\lambda }g_{\mu \nu }=-\alpha \omega _{\lambda }g_{\mu \nu
}\equiv Q_{\lambda \mu \nu }, \tilde{\nabla}_{\lambda }g^{\mu \nu }=\alpha
\omega _{\lambda }g^{\mu \nu }.
\end{equation}%
where the constant $\alpha $ denotes the Weyl gauge coupling constant. The Weyl connection $\tilde{\Gamma}%
_{\mu \nu }^{\lambda }$ can be obtained from the nonmetricity
condition (\ref{nmetr}) as
\begin{equation}  \label{Con}
\tilde{\Gamma}_{\mu \nu }^{\lambda }=\Gamma _{\mu \nu }^{\lambda }+\frac{1}{2%
}\alpha \Big[\delta _{\mu }^{\lambda }\omega _{\nu }+\delta _{\nu }^{\lambda
}\omega _{\mu }-g_{\mu \nu }\omega ^{\lambda }\Big]=\Gamma _{\mu \nu
}^{\lambda }+\Psi _{\mu \nu }^{\lambda },
\end{equation}%
where $\Gamma _{\mu \nu }^{\lambda }$ is the Levi-Civita connection
of the metric $g_{\mu \nu }$
\begin{equation}
\Gamma _{\mu \nu }^{\lambda }=\frac{1}{2}g^{\lambda \sigma}\left(\partial
_\nu g_{\sigma \mu}+\partial_\mu g_{\sigma \nu}-\partial _\sigma g_{\mu
\nu}\right).
\end{equation}

In the following the geometrical and physical quantities in
Weyl geometry will be denoted by a tilde. From the contraction of Eq.~(\ref{Con}) we obtain
\begin{equation}
\omega _{\mu }=\frac{1}{2\alpha }\left( \tilde{\Gamma}_{\mu }-\Gamma _{\mu
}\right) ,
\end{equation}
which gives the geometrical interpretation of the Weyl vector as the difference of the Weyl and Levi-Civita connections.

A geometrical quantity playing an important role in many applications is the field strength $%
F_{\mu\nu}$ of the Weyl vector $\omega_\mu$, is defined according to,
\begin{equation}  \label{W}
\tilde{F}_{\mu\nu} = \tilde{\nabla}_{[\mu} \omega_{\nu]} = \nabla_{[\mu}
\omega_{\nu]} = \partial_{[\mu} \omega_{\nu]}=\partial _{\mu}\omega
_\nu-\partial _\nu \omega _\mu.
\end{equation}

The action of the covariant
derivative commutators on vectors and covectors is expressed as
\begin{subequations}
\begin{align}
\left( \tilde{\nabla}_{\mu }\tilde{\nabla}_{\nu }-\tilde{\nabla}_{\nu }%
\tilde{\nabla}_{\mu }\right) v^{\sigma }& =\tilde{R}\indices{^\sigma_{\rho
\mu \nu}}v^{\rho }, \\
\left( \tilde{\nabla}_{\mu }\tilde{\nabla}_{\nu }-\tilde{\nabla}_{\nu }%
\tilde{\nabla}_{\mu }\right) v_{\sigma }& =-\tilde{R}\indices{^\rho_{\sigma
\mu \nu}}v_{\rho }.
\end{align}%
where we have introduced the Weyl curvature tensor $\tilde{R}_{\mu \nu \sigma }^{\lambda }$ defined according to the definition
\end{subequations}
\begin{equation}
\tilde{R}_{\mu \nu \sigma }^{\lambda }=\partial _{\nu }\tilde{\Gamma}_{\mu
\sigma }^{\lambda }-\partial _{\sigma }\tilde{\Gamma}_{\mu \nu }^{\lambda }+%
\tilde{\Gamma}_{\rho \nu }^{\lambda }\tilde{\Gamma}_{\mu \sigma }^{\rho }-%
\tilde{\Gamma}_{\rho \sigma }^{\lambda }\tilde{\Gamma}_{\mu \nu }^{\rho },
\end{equation}

The Weyl curvature tensor has the symmetry properties
\begin{align}
& \tilde{R}_{\mu \nu \rho \sigma }=-\tilde{R}_{\mu \nu \sigma \rho }, \\
& \tilde{R}_{\mu \nu \rho \sigma }=-\tilde{R}_{\nu \mu \rho \sigma }+\alpha
g_{\mu \nu }F_{\rho \sigma }, \\
& \tilde{R}_{\mu \nu \rho \sigma }=\tilde{R}_{\rho \sigma \mu \nu }+\frac{%
\alpha }{2}\Big(g_{\mu \nu }F_{\rho \sigma }-g_{\rho \sigma }F_{\mu \nu }
\notag \\
& \quad +g_{\nu \sigma }F_{\mu \rho }-g_{\nu \rho }F_{\mu \sigma }+g_{\mu
\rho }F_{\nu \sigma }-g_{\mu \sigma }F_{\nu \rho }\Big), \\
& \tilde{R}_{\mu \nu }=\tilde{R}_{\nu \mu }+2\alpha F_{\mu \nu }.
\end{align}

The first and the second contractions of the Weyl curvature tensor are given by
\begin{equation}
\tilde{R}_{\mu \nu }=\tilde{R}_{\mu \lambda \nu }^{\lambda },\tilde{R}%
=g^{\mu \sigma }\tilde{R}_{\mu \sigma }.
\end{equation}%
Thus we obtain for the Weyl scalar $\tilde{R}$ the expression
\begin{equation}
\tilde{R}=R-3\alpha \nabla _{\mu }\omega ^{\mu }-\frac{3}{2}\left( \alpha
\right) ^{2}\omega _{\mu }\omega ^{\mu },  \label{R}
\end{equation}%
where $R$ is the Ricci scalar obtained with the help of the Levi-Civita connection of the Riemannian geometry.

 With respect to a conformal transformation with a conformal factor $\Sigma (x)$, the variations of the
metric tensor, of the Weyl field, and of a scalar field $\phi $ are given by
\begin{eqnarray}  \label{a2}
\tilde g_{\mu\nu} = \Sigma ^2(x)g_{\mu\nu}, \tilde{ \omega} _\mu = \omega_\mu - \frac{%
2}{\alpha} \partial_\mu \ln\Sigma (x), \tilde{ \phi} = \Sigma^{-1} (x) \phi. \nonumber\\
\end{eqnarray}

\subsection{Weyl geometric gravity-action and field equations}

The simplest gravitational Lagrangian density in Weyl geometry, which is conformally
invariant, was first considered by Weyl \cite{Weyl1, Weyl2, Weyl3}, and can be defined according to
\cite{Gh3a,Gh4a,Gh5a,Gh6a,Gh7a}
\begin{eqnarray}  \label{inA}
\tilde{L}_{Weyl}=\Big[\, \frac{1}{4!}\,\frac{1}{\xi^2}\,\tilde R^2 - \frac14\,
\tilde{F}_{\mu\nu}^{\,2} \Big]\sqrt{-g},
\end{eqnarray}
where  $\xi < 1$ denotes the parameter of the perturbative coupling. We will linearize the Lagrangian $\tilde{L}_{Weyl}$ in the curvature by introducing the scalar field $\phi_0$ according to the definition \cite{Gh3a,Gh4a,Gh5a,Gh6a,Gh7a}
\begin{equation}
\tilde{R}^2\rightarrow 2 \phi_0^2\,\tilde R-\phi_0^4,
\end{equation}
The
new Lagrangian density obtained after this substitution is equivalent mathematically to the original one. This result follows from the use of the solution of the equation of motion of $\phi_0$, $\phi_0^2=%
\tilde {R}$ in the new Lagrangian $\tilde{L}_{Weyl}$.
Hence, via this substitution, we obtain a new geometric Lagrangian, defined in Weyl geometry, which includes a scalar degree
of freedom, and is given by
\begin{eqnarray}\label{alt3}
\tilde{L}_{Weyl}=\sqrt{-g} \Big[\frac{1}{12}\frac{1}{\xi^2}\,\phi_0^2\,\tilde R
-\frac14 \,\tilde{F}_{\mu\nu}^2-\frac{\phi_0^4}{4!\,\xi^2} \Big].
\end{eqnarray}

The Lagrangian (\ref{alt3}) gives the simplest gravitational Lagrangian
density fully including the Weyl gauge symmetry, since it is conformally invariant. $\tilde{L}_{Weyl}$ contains a spontaneous breaking of the conformal symmetry,
which leads to an Einstein-Proca Lagrangian for $\omega _\mu$.

To obtain the gravitational action of Weyl geometric gravity we substitute in Eq.~(\ref{alt3}) $\tilde{R}$ by its expression given by Eq.~(\ref%
{R}). Then,  after a gauge transformation, and by redefining the
geometrical and physical quantities, we find the Riemann space action of Weyl geometric gravity, given by \cite{Gh3a,Gh4a,Gh5a}
\begin{eqnarray}  \label{a3}
\mathcal{S }_{Weyl}&=& \int d^4x \sqrt{-g} \Bigg[ \frac{1}{12} \frac{\phi^2}{\xi^2} %
\Big( R - 3\alpha\nabla_\mu \omega^\mu - \frac{3}{2} \alpha^2 \omega_\mu
\omega^\mu \Big)  \notag \\
&&- \frac{1}{4!}\frac{\phi^4}{\xi^2} - \frac{1}{4} \tilde{F}_{\mu\nu} \tilde{%
F}^{\mu\nu} \Bigg],
\end{eqnarray}
The action $S_{Weyl}$ is fully invariant under conformal transformations.

By varying the action (%
\ref{a3}) with respect to the metric tensor we obtain the field equations of Weyl geometric gravity as \cite%
{I3,I4}, 
\begin{eqnarray}  \label{b2a}
&&\frac{\phi ^{2}}{\xi ^{2}}\Big(R_{\mu \nu }-\frac{1}{2}Rg_{\mu \nu }\Big)+%
\frac{1}{\xi ^{2}}\Big(g_{\mu \nu }\Box -\nabla _{\mu }\nabla _{\nu }\Big)%
\phi ^{2}  \notag \\
&&-\frac{3\alpha }{2\xi ^{2}}\Big(\omega ^{\rho }\nabla _{\rho }\phi
^{2}g_{\mu \nu }-\omega _{\nu }\nabla _{\mu }\phi ^{2}-\omega _{\mu }\nabla
_{\nu }\phi ^{2}\Big)  \notag \\
&&+\frac{3\alpha ^{2}}{4\xi ^{2}}\phi ^{2}\Big(\omega _{\rho }\omega ^{\rho
}g_{\mu \nu }-2\omega _{\mu }\omega _{\nu }\Big) -6\tilde{F}_{\rho \mu }%
\tilde{F}_{\sigma \nu }g^{\rho \sigma }  \notag \\
&&+\frac{3}{2}\tilde{F}_{\rho \sigma }^{2}g_{\mu \nu }+\frac{1}{4\xi ^{2}}%
\phi ^{4}g_{\mu \nu }=0.
\end{eqnarray}

The trace of Eq.~(\ref{b2a}) gives
\begin{equation}  \label{b3n}
\Phi R+3\alpha \omega ^{\rho }\nabla _{\rho }\Phi -\Phi ^{2}-\frac{3}{2}%
\alpha ^{2}\Phi \omega _{\rho }\omega ^{\rho }-3\Box \Phi =0,
\end{equation}%
where we have denoted $\Phi \equiv \phi ^{2}$. The variation of
the action (\ref{a3}) with respect to the scalar field $\phi $ is given by
\begin{equation}  \label{b4}
R-3\alpha \nabla _{\rho }\omega ^{\rho }-\frac{3}{2}\alpha ^{2}\omega _{\rho
}\omega ^{\rho }-\Phi =0,
\end{equation}%
This relation  gives the equation of motion of the scalar field $%
\phi $. From Eqs.~(\ref{b3n}) and (\ref{b4}) we find
\begin{equation}  \label{b5}
\Box \Phi-\alpha \nabla _{\rho }(\Phi\omega ^{\rho })=0.
\end{equation}

For the Weyl vector we obtain the equation of motion
\begin{equation}  \label{Fmunu}
4\xi ^{2}\nabla _{\nu }\tilde{W}^{\mu \nu }+\alpha ^{2}\Phi\omega ^{\mu
}-\alpha \nabla ^{\mu }\Phi=0.
\end{equation}

The application to both sides of Eq.~(\ref{Fmunu}) of the operator $\nabla _{\mu
} $ leads to Eq.~\eqref{b5}. Thus, the field equations of the theory are consistent.

\section{Exact static analytical solution and the tangential rotation curves in Weyl geometric gravity}\label{sect2}

In the present Section we introduce an exact static, spherically symmetric solution of the field equations of the Weyl geometric gravity, obtained in a specific gauge in which the Weyl vector field has only a spacelike component \cite{I3,I4}. The expression of the tangential velocity of massive test particles moving in this geometry is also presented.

\subsection{Exact solution of the Weyl geometric gravity field equations}

In the following we will look for a static, spherically symmetric solution of the Weyl geometric gravity field equations. To this purpose we adopt a static and spherically symmetric geometry, and we introduce the set of coordinates  $\left( ct,r,\theta ,\varphi
\right) $ on the base manifold. For the spacetime interval we adopt the expression \cite{LaLi}
\begin{equation}
ds^{2}=e^{\nu (r)}c^{2}dt^{2}-e^{\lambda (r)}dr^{2}-r^{2}\left(d\theta ^{2}+\sin ^{2}\theta d\varphi
^{2}\right),
\label{line}
\end{equation}%
where the metric coefficients are functions of the radial coordinate $r$ only. We suppose that the Weyl vector $\omega _\mu$ has only
one non-vanishing component, and thus
\be
\omega _{\mu }=(0,\omega _{1}(r),0,0).
\ee

For this form of $\omega
_{\mu } $ the Weyl field strength tensor vanishes identically,  $\tilde{F}_{\mu \nu }\equiv 0$.  Eq.~(\ref{Fmunu}) immediately gives
\begin{equation}
\Phi ^{\prime }=\alpha \omega _1\Phi .  \label{15}
\end{equation}

For the full set of the static spherically symmetric field equations of Weyl geometric gravity we refer the reader to Refs. \cite{I3,I4}.
By assuming that the metric tensor components satisfy the condition $\nu +\lambda =0$, the scalar field equation takes the form
\begin{equation}
\Phi ^{\prime \prime }-\frac{3}{2}\frac{\Phi ^{\prime 2}}{\Phi ^{2}}=0,
\end{equation}%
and it has  the general solution
\begin{equation}
\Phi (r)=\frac{C_{1}}{C_2^2}\frac{1}{\left( 1+\frac{r}{C_{2}}\right) ^{2}},  \label{40}
\end{equation}%
where $C_{1}$ and $C_{2}$ are arbitrary constants of integration. The
radial component of the Weyl vector is given by
\begin{equation}\label{omeg}
\omega _{1}(r)=\frac{1}{\alpha }\frac{\Phi ^{\prime }(r)}{\Phi (r)}=-\frac{2}{\alpha C_2}%
\frac{1}{1+\frac{r}{C_{2}}},
\end{equation}

With the use of the above forms of the Weyl vector and of the scalar field, the field
equations of the Weyl geometric gravity theory possess an exact solution, given by \cite{I3,I4},
\begin{eqnarray}
e^{-\lambda (r)} &=&e^{\nu (r)}=\frac{r\left( 12C_{3}C_{2}^{2}-C_{1}-4\right) }{%
4C_{2}}\nonumber\\
&&+\frac{1}{4}\left( 12C_{3}C_{2}^{2}-C_{1}-8\right)  \notag  \label{58}
\\
&&+\frac{C_{2}}{12}\left[ 12C_{3}C_{2}^{2}-C_{1}-12\right] \frac{1}{r}%
+C_{3}r^{2},
\end{eqnarray}%
where $C_{3}$ is another arbitrary constant of  integration.

The solution depends on three arbitrary independent integration. Depending on their parametrization,
 the exact solution (\ref{58}) can be represented in several distinct ways
forms. If we choose the constants so that they obey the condition, $12C_{3}C_{2}^{2}-C_{1}-8=4$%
, or $C_{3}C_{2}^{3}-C_{1}C_{2}/12=C_{2}$, then the line element (\ref%
{58}) becomes \cite{I3,I4}
\begin{equation}
e^{-\lambda (r)}=e^{\nu (r)}=1+\frac{2}{C_{2}}r+C_{3}r^{2},
\end{equation}%
thus representing an extension of the
de Sitter static, cosmological metric. From its form it follows that this metric is not asymptotically
flat.

The parametrization of the metric (\ref{58}) we will consider in the following is obtained
by introducing a new variable $r_g$, according to the definition
\begin{equation}
\frac{C_{2}}{12}\left[ 12C_{3}C_{2}^{2}-C_{1}-12\right] =-r_{g}.
\end{equation}%

We interpret $r_{g}=2GM/c^{2}$ from a physical point of view as the gravitational
radius of an object of mass $M$. Therefore, we obtain the following relations between the integration constants
\be
12C_{3}C_{2}^{2}-C_{1}-8=4\left( 1-3r_{g}/C_{2}\right) ,
\ee
and
\be
12C_{3}C_{2}^{2}-C_{1}-4=4\left( 2-3r_{g}/C_{2}\right),
\ee
respectively. For the integration constant $C_{3}$ we obtain the expression
\begin{equation}
C_{3}=\frac{1}{C_{2}^{2}}\left( 1+\frac{C_{1}}{12}-\frac{r_{g}}{C_{2}}%
\right) .
\end{equation}

Therefore, with this parametrization of the integration constants,  the metric (\ref{58}) becomes
\begin{eqnarray}
e^{\nu (r) }=e^{-\lambda (r) }&=&1-\frac{3r_{g}}{C_{2}}-\frac{r_{g}}{r}+\left( 2-3%
\frac{r_{g}}{C_{2}}\right) \frac{r}{C_{2}}  \notag \\
&&+\left( 1+\frac{C_{1}}{12}-\frac{r_{g}}{C_{2}}\right) \frac{r^{2}}{%
C_{2}^{2}}.  \label{mfin}
\end{eqnarray}

From Eq.~(\ref{omeg}) we obtain the constant $C_2$ as
\begin{equation}
C_2=-\frac{\alpha}{2}\omega _1(r)-r.
\end{equation}
From the expression of the scalar field (\ref{40}) we find the relation
\begin{equation}
C_1=\frac{4}{\alpha ^2}\frac{\Phi(r)}{\omega _1^2}.
\end{equation}

Note that a black hole solution similar to Eq.~(\ref{mfin})
was found in Weyl conformal gravity \cite{M1}, and in the dRGT massive
gravity theory \cite{Pi}.

\subsection{Rotational velocities in static spherically symmetric geometries}

The equations of motion of a massive particle in the gravitational field described by the general spherically symmetric metric (\ref{line}) can be obtained from the Lagrangian \cite{LaLi, Lake},
\begin{equation}
\mathcal{L}_W=\left[ e^{\nu \left( r\right) }\left( \frac{cdt}{ds}\right)
^{2}-e^{\lambda \left( r\right) }\left( \frac{dr}{ds}\right)
^{2}-r^{2}\left( \frac{d\Omega }{ds}\right) ^{2}\right] ,
\end{equation}

By considering the motion restricted to the galactic plane with $\theta =\pi /2$ we obtain $d\Omega ^{2}=d\phi ^{2}$. From the Lagrange equations it follows the existence  of two constants of motion, the energy of the particle $E$, and its angular momentum $l$, respectively. Their expressions are given by $E=e^{\nu
(r)}c^{3}\dot{t}$ and $\ell=cr^{2}\dot{\phi}$, respectively. Due to the normalization of the four-velocity as $u^{\mu }u_{\mu }=1$, one can easily find
the constraint $1=e^{\nu \left( r\right) }c^{2}\dot{t}^{2}-e^{\lambda (r)}\dot{r}%
^{2}-r^{2}\dot{\phi}^{2}$, from which, with the use of the constants of motion, one finds
finds the conserved energy of the particle as
\begin{equation}
\frac{E^{2}}{c^{2}}=e^{\nu +\lambda }\dot{r}^{2}+e^{\nu }\left(1+ \frac{\ell^{2}%
}{c^{2}r^{2}}\right) .  \label{energy}
\end{equation}

We interpret Eq.~(\ref{energy}) as corresponding to the radial displacement of a
massive particle in Newtonian mechanics.  The particle has a velocity $\dot{r}
$, a position dependent effective mass $m_{\mathrm{eff}}(r)=2e^{\nu (r) +\lambda (r)}$,  and, of course,  a conserved energy $E^{2}$.
The motion takes place in the presence of an effective potential $V_{\mathrm{%
eff}}\left( r\right) $, which can be obtained as
\begin{equation}\label{Veff}
V_{\mathrm{eff}}\left( r\right) =e^{\nu (r)}\left(1+ \frac{\ell^{2}}{c^{2}r^{2}}%
\right) .
\end{equation}

For stable circular particle orbits, satisfying the conditions  $\dot{r}=0$ and $\partial V_{%
\mathrm{eff}}/\partial r=0$, respectively, the conserved
energy and angular momentum become
\begin{equation}
\frac{E^{2}}{c^{4}}=\frac{e^{\nu }}{1-r\nu ^{\prime }/2},\;\;
\ell^{2}=\frac{c^{2}}{2}\frac{r^{3}\nu ^{\prime }}{1-r\nu ^{\prime }/2},
\label{cons1}
\end{equation}

The spatial velocity $v$ of the massive particle is obtained as \cite%
{LaLi}
\begin{equation}
v^{2}(r)=e^{-\nu }\left[ e^{\lambda }\left( \frac{dr}{dt}\right)
^{2}+r^{2}\left( \frac{d\Omega }{dt}\right) ^{2}\right] .
\end{equation}%

For $\dot{r}=0$ the tangential velocity of a massive test particles
has the  expression
\begin{equation}
v_{tg}^{2}(r)=e^{-\nu }r^{2}\left( \frac{d\Omega }{dt}\right) ^{2}=e^{-\nu
}r^{2}c^{2}\left( \frac{d\Omega }{ds}\right) ^{2}\left( \frac{ds}{cdt}%
^{2}\right) .
\end{equation}

For a motion in the equatorial plane with $\theta =\pi /2$, we find for the tangential velocity the simple expressions
\begin{equation}
\frac{v_{tg}^{2}(r)}{c^{2}}=c^{2}\frac{e^{\nu }}{r^{2}}\frac{\ell^{2}}{E^{2}}\,,
\end{equation}%
and
\begin{equation}
\frac{v_{tg}^{2}(r)}{c^{2}}=\frac{r\nu ^{\prime }}{2},  \label{vtg2}
\end{equation}%
respectively.

By using the above expression of the tangential velocity, the angular momentum of the particle can be written as
\be
\ell ^2=r^2\frac{v_{tg}^2(r)}{1-v_{tg}^2(r)/c^2}.
\ee

For the effective potential we obtain
\be\label{pot}
V_{eff}(r)=\frac{e^{\nu (r)}}{1-v_{tg}^2(r)/c^2}.
\ee

If the tangential velocity of massive test particles can be found from
observational data, or by using some theoretical models, one can obtain the metric tensor component $\nu (r)$
in the dark matter dominated region of a galaxy  as
\begin{equation}
\nu (r)=2\int {\frac{v_{tg}^{2}(r)}{c^{2}}\frac{dr}{r}}.  \label{metricnu}
\end{equation}

\subsection{Tangential velocity in Weyl geometric gravity}

In the case of the static spherically symmetric exact solution of Weyl geometric
gravity, with line element given by Eq.~(\ref{mfin}) we have
\begin{equation}
\nu ^{\prime }=\frac{\frac{r_{g}}{r^{2}}+\left( 2-3\frac{r_{g}}{C_{2}}%
\right) \frac{1}{C_{2}}+2\left( 1+\frac{C_{1}}{12}-\frac{r_{g}}{C_{2}}%
\right) \frac{r}{C_{2}^{2}}}{1-\frac{3r_{g}}{C_{2}}-\frac{r_{g}}{r}+\left(
2-3\frac{r_{g}}{C_{2}}\right) \frac{r}{C_{2}}+\left( 1+\frac{C_{1}}{12}-%
\frac{r_{g}}{C_{2}}\right) \frac{r^{2}}{C_{2}^{2}}},
\end{equation}%
giving for the tangential velocity the expression

\begin{equation}
\frac{2v_{tg}^{2}}{c^2}=\frac{\frac{r_{g}}{r}+\left( 2-3\frac{r_{g}}{C_{2}}%
\right) \frac{r}{C_{2}}+2\left( 1+\frac{C_{1}}{12}-\frac{r_{g}}{C_{2}}%
\right) \frac{r^{2}}{C_{2}^{2}}}{1-\frac{3r_{g}}{C_{2}}-\frac{r_{g}}{r}%
+\left( 2-3\frac{r_{g}}{C_{2}}\right) \frac{r}{C_{2}}+\left( 1+\frac{C_{1}}{%
12}-\frac{r_{g}}{C_{2}}\right) \frac{r^{2}}{C_{2}^{2}}},
\end{equation}%
or,
\begin{equation}\label{vtg}
v_{tg}=c\sqrt{\frac{r_{g}}{2r}}\sqrt{\frac{1+2\left( 1-\frac{3}{2}\frac{r_{g}}{C_{2}}%
\right) \frac{r^{2}}{r_{g}C_{2}}+2\left( 1+\frac{C_{1}}{12}-\frac{r_{g}}{%
C_{2}}\right) \frac{r^{3}}{r_{g}C_{2}^{2}}}{1-\frac{3r_{g}}{C_{2}}-\frac{%
r_{g}}{r}+2\left( 1-\frac{3}{2}\frac{r_{g}}{C_{2}}\right) \frac{r}{C_{2}}+\left( 1+%
\frac{C_{1}}{12}-\frac{r_{g}}{C_{2}}\right) \frac{r^{2}}{C_{2}^{2}}}}.
\end{equation}

The first term in the above equation is nothing but the Keplerian velocity
of the particle, $v_{K}=\sqrt{GM/r}$. The second term gives the Weyl
geometrical corrections to the ordinary Newtonian velocity.

In the following we assume that $M$ represents the baryonic mass of the galaxy.
For a baryonic mass of the order of $M=10^{10}M_{\odot}$, $r_{g}=2GM/c^{2}$
has the value $r_{g}=2.96\times 10^{15}$ cm. Taking into account that 1 kpc $%
=3.08\times 10^{21}$ cm, and that 1 cm $=1/3.08\times 10^{21}$ kpc, it
follows that $r_{g}=9.61\times 10^{-7}\approx 10^{-6}$ kpc. However, one can
assume for $r_g$ a range of $r_g\in \left(10^{-8}, 10^{-4}\right)$ kpc.

The
constant $C_2$ has the physical dimensions of length, and it can take both positive and
negative values. If $C_2>0$, it must satisfy the condition $C_2>>3r_g$,
so that the term $1-3r_g/C_2$ in the metric is always positive, and tends to
1 in the Newtonian limit. There are no similar restrictions for negative
values of $C_2$. $C_2$ is related to the behavior of the Weyl vector field.

On the other hand, $C_1$ is a dimensionless constant, describing the magnitude
of the auxiliary scalar field. Similarly to $C_2$, it can take both positive and negative
numerical values. In order to avoid any singular or unphysical behavior in the metric
for small $r$, we assume that the range of the radial coordinates is
restricted to values $r>r_g$.

In the limit $\left|C_2\right|\rightarrow \infty$, we reobtain from Eq. (\ref%
{vtg}) the general relativistic limit of the tangential velocity,
\begin{equation}
\lim_{\left|C_2\right|\rightarrow \infty}v_{tg}=c\sqrt{\frac{r_g}{2r}}%
\left(1-\frac{r_g}{r}\right)^{-1/2},
\end{equation}
corresponding to the Schwarzschild metric with $e^\nu=1-r_g/r$. In the limiting case $r\rightarrow \infty$, we have
$v_{tg}\rightarrow c$, that is, the velocity of the test particles tend to the speed of light. However, if the integration constants satisfy the condition $1+C_1/12-r_g/C_2=0$, in the large $r$ limit the tangential velocity tends to $\lim_{r\rightarrow \infty}v_{tg}=c/\sqrt{2}$.

For large values of $r$, the tangential velocity behaves as
\bea
\frac{v_{tg}^2}{c^2}&\approx& 1+\frac{6 C_2 (3 r_g-2 C_2)}{r \left[(C_1+12) C_2-12
   r_g\right]}\nonumber\\
   &&+\frac{12 C_2^2 \left[-(C_1-12) C_2^2+3 (C_1-8)
   C_2 r_g+18 r_g^2\right]}{r^2 \left[(C_1+12) C_2-12
   r_g\right]^2}\nonumber\\
  && +O\left(\left(\frac{1}{r}\right)^3\right),
\eea
which also show that in the limit of large $r$ the tangential velocity tends towards the speed of light.

\section{Fitting the Weyl geometric rotation curves with the SPARC sample}\label{sect3}

In the present Section we consider a detailed comparison of the theoretical predictions of the tangential velocities obtained from the exact solution of Weyl geometric gravity with the observational data. For this comparison we use the data of the SPARC sample, which contains the rotation curves of a large number of galaxies,

We consider galaxies both with and without bulge velocities, and we perform the comparison with all galaxies having an acceptable number of observational points.

\subsection{The SPARC dataset}

\begin{figure}[htbp]
\centering
\includegraphics[width=6.5cm]{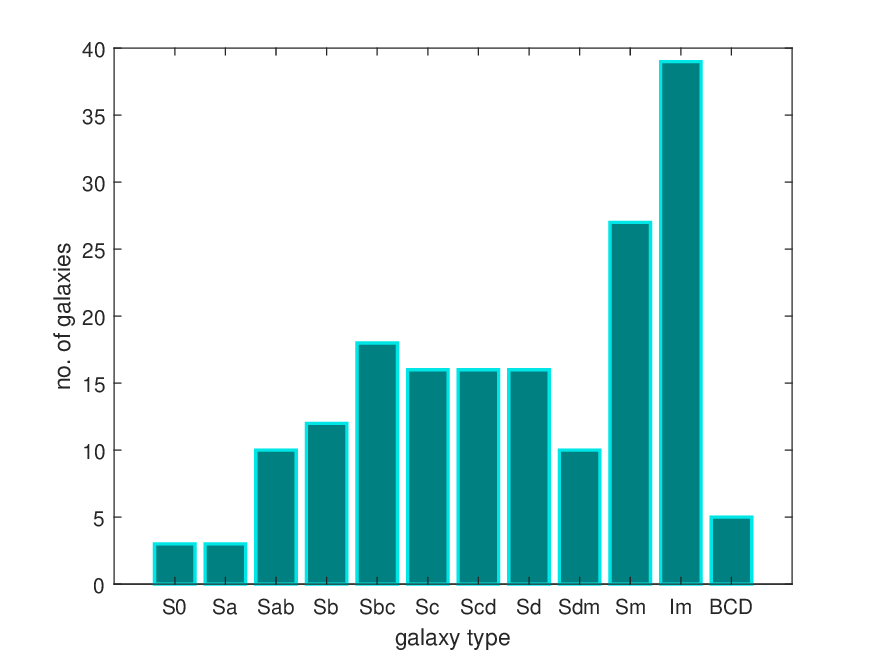}
\caption{Distribution of the galaxies of the SPARC sample according to their
morphological type.}
\label{fig1}
\end{figure}

The SPARC sample consists of the rotation curves data of 175 galaxies \cite%
{S1, S2, S3}. In the sample all rotationally supported morphological galactic types are included. The distribution of the galaxies of the SPARC sample
according to their morphological types are presented in Fig.~\ref{fig1}. As one can see from Fig.~\ref{fig1}), the SPARC sample contains data on a large variety of galaxy types.

A large number of astrophysical/astronomical data are also provided in the SPARC sample,
including the Hubble types, distance, inclination, total luminosity,
effective radius, and effective surface brightness at [3.6], total HI mass,
and the asymptotically flat rotation velocity, respectively.

\begin{figure*}[htbp]
\centering
\includegraphics[width=6.5cm]{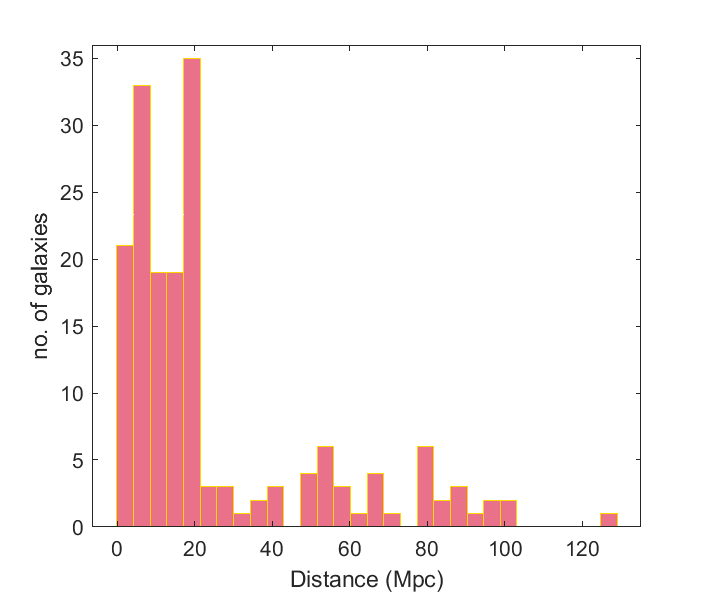} %
\includegraphics[width=6.5cm]{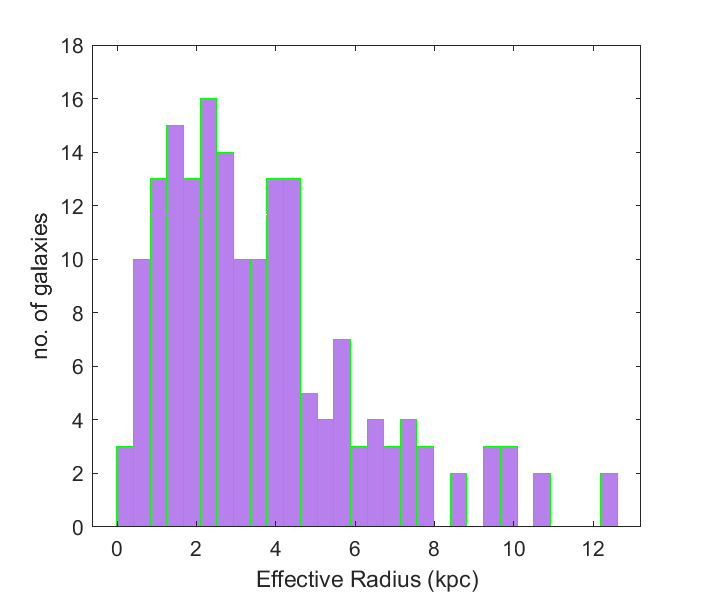}
\caption{Distribution of the galactic distances (left panel) and of the
effective radius (right panel) in the SPARC sample.}
\label{fig2}
\end{figure*}

The distance and radius
distributions of the galaxies in the sample are shown in Fig.~\ref{fig2}.

Data on the distribution of the stellar masses, together with 21 cm
observations, tracing the atomic gas, are also added to the sample. The rotation curves are
obtained from the 21 cm velocity fields \cite{S1,S2,S3}, and information on the observations of the ionized
interstellar medium are also provided.

Up to now, the SPARC sample is the
largest existing galactic database, giving for every galaxy not only the rotation
curves, but also spatially resolved data, showing the distribution of both
stars and gas \cite{S1,S2,S3}.

\begin{figure*}[htbp]
\centering
\includegraphics[width=6.5cm]{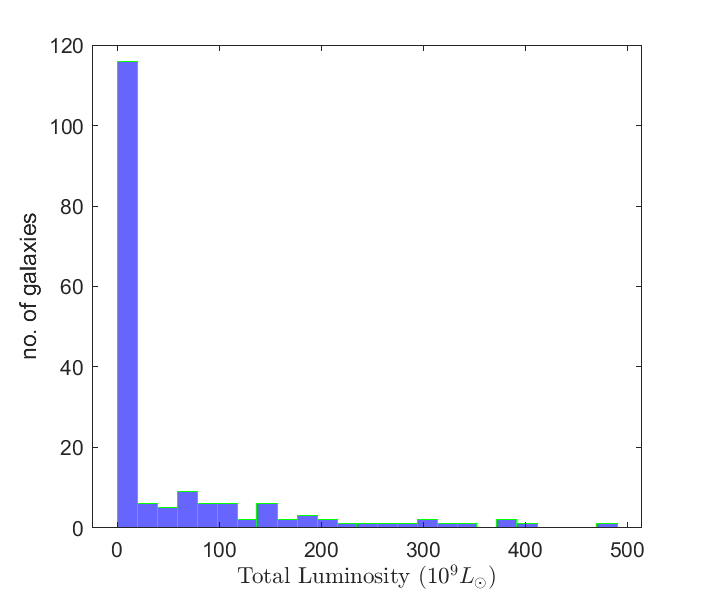}
\includegraphics[width=6.5cm]{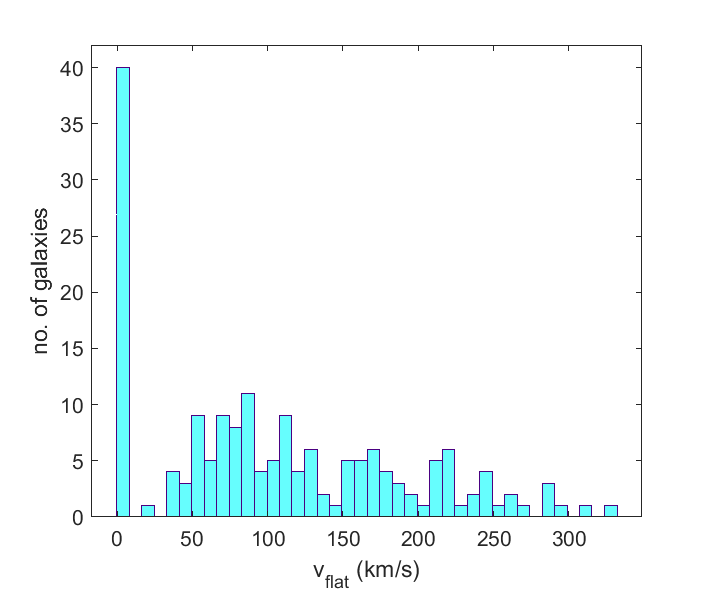}
\caption{Distribution of the galactic luminosities (left panel) and of the Asymptotically Flat Rotation Velocities of the galaxies (right panel) in the
SPARC dataset.}
\label{fig3}
\end{figure*}

The SPARC sample contains galaxies with rotation velocities in the range $20
< Vf < 300$ km/ s, and luminosities in the interval $10^7 <L_{[3.6]} < 5
\times 10^{11} L_{\odot}$.  The distribution of the galactic luminosities and of the Asymptotically Flat Rotation Velocities of
the galaxies in the SPARC dataset is presented in Fig.~\ref{fig3}.

\begin{figure*}[htbp]
\centering
\includegraphics[width=6.5cm]{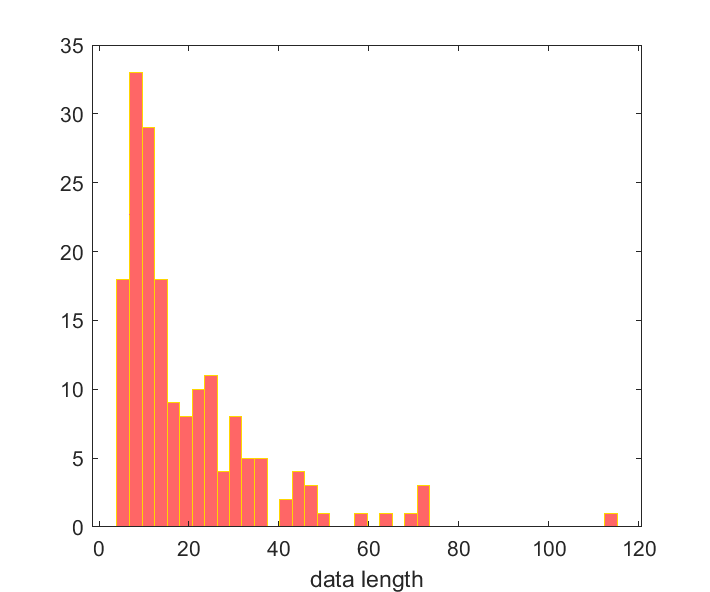} %
\includegraphics[width=6.5cm]{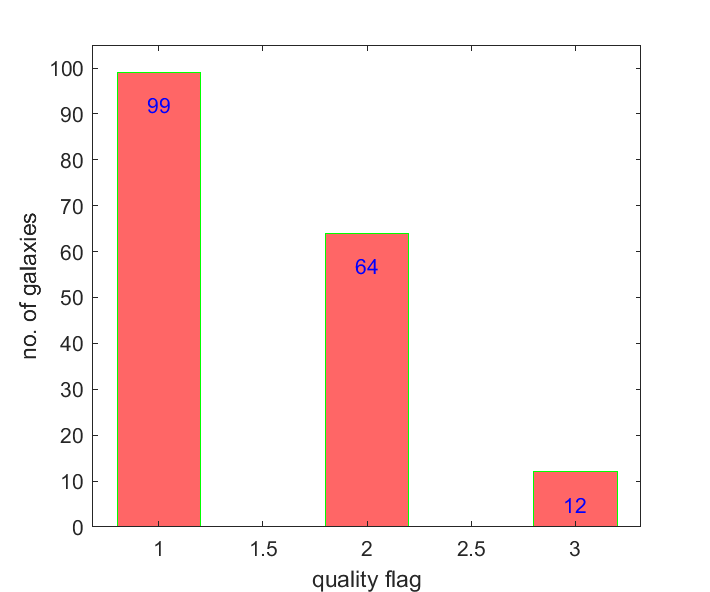}
\caption{Distribution of the number of the observational points (left panel)
and of the quality flag (right panel) in the SPARC sample.}
\label{fig4}
\end{figure*}

The distribution of the number of data points for the observations of the
galactic rotation curves, as well as their quality flags, are presented in
Fig.~\ref{fig4}.

The SPARC dataset contains significant information not only about large individual galaxies, but as
well as about small ones. Moreover, low surface brightness and low mass
galaxies are also well represented in SPARC \cite{S1,S2,S3}. This represents a significant
difference as compared to flux selected samples, which contain data only on the
interval $M_{*} > 10^9 M_{\odot}$, and $V_f > 100$ km/s, respectively \cite%
{S3}.

\subsection{Material and methods}\label{secMM}

To test the validity of the expression of the velocity of the test particles
in the Weyl geometric gravity theory, as given by Eq. (\ref{vtg}), we have
fitted the predicted theoretical velocity profile with the total
observational velocity of the particles that can be obtained from the SPARC
database. In the following we will present first the theoretical model used for obtaining the fits.

\subsubsection{The theoretical model}

We assume that the total velocity $v_{t}$ can be obtained from the relation \cite{S3}
\begin{equation}
v_{t}=\sqrt{v_{g}|v_{g}|\!+\!\Upsilon _{d}\!\times
\!v_{d}|v_{d}|\!+\!\Upsilon _{b}\!\times \!v_{b}|v_{b}|\!+\!v_{tg}^{2}},
\label{v_tot}
\end{equation}%
where $v_{t}$ is the total velocity, including the contributions of both
baryonic and dark matter, and $v_{g}$, $v_{d}$, $v_{b}$ and $v_{tg}$ denote
the contributions from the velocity of the gas, of the disk, of the bulge,
and of the Weyl geometric gravity effects, respectively. $\!\Upsilon _{d}\!$
and $\!\Upsilon _{b}\!$ denote the stellar mass-to-light ratios for the disk
and the stellar bulge, respectively.

\subsubsection{The fitting procedure}

To perform the fitting, and by taking into account the observational errors,
we have looked for the minimum of the objective function
\begin{equation}
\chi^2(r_g,C_1,C_2,\Upsilon_d,\Upsilon_b)=\frac{1}{n-k}\sum_{i=1}^{n}%
\frac{(v_{t,i}-v_{obs,i})^{2}}{\sigma _{i}^{2}},  \label{chi2}
\end{equation}
where $n$ is the length of data, and $k$ is the number of parameters that
must be estimated, that is, $k=4$ for bulgeless galaxies, and $k=5$ for
galaxies with bulge velocity data, respectively.
We selected the galaxies with length data larger than the number of parameters to be optimized (4, respectively 5).
We used  Multi Start and Global Search methods (see \cite{ugray2007} for a full description of the algorithm and the scatter-search method of generating trial points).
For both methods the solver attempts to find multiple local solutions to an optimization problem by starting from various points.
Global Search analyzes start points, discarding ones that are improbable to improve the current best local minimum. On the other hand,  Multi Start runs the local solver on all start points (or, optionally, all those that meet feasibility criteria concerning bounds or inequality constraints.).
In the present investigation the number of start points was set to 10000.

The intervals for the parameters appearing in the tangential velocity were chosen as follows: $r_g \in [10^{-23},\   10^{-2}] $,  $C_1 \in [-10^{11},\  10^{14}] $,  $C_2 \in [-10^{11}, \ 10^{14}] $. We also imposed the condition $ r_g/C_2<1 $, and we have limited the values of $\Upsilon_d$ and $\Upsilon_b$ to the interval [0.1,5].

\subsection{Fitting results}

In the following we consider independently the two group of galaxies without and with  bulge velocity information.  The number of fitting parameters is different in the two cases (4 respective 5).

\subsection{Fitting results for galaxies without bulge velocity information}

As a first example of the comparison of the theoretical model of the rotation curves as given by Weyl geometric gravity and observations we consider the fitting of the galaxies of the SPARC sample that do not have a bulge velocity component. The number of galaxies we have considered in our analysis is 76. The comparison of the observational data and of the theoretical model are presented, for a selected sample of 55 galaxies, in Figs.~\ref{fig_vit1}, \ref{fig_vit2}, and \ref{fig_vit3}, respectively.

In each plot we have presented,  for each galaxy, the observational data with their error bars, the contribution of the baryonic matter, the contribution coming from the Weyl geometric gravity, as well as the total velocity, as given by Eq.~(\ref{v_tot}).

The numerical values of the model parameters $\left(r_g,C_1,C_2\right)$, obtained from the fitting, are presented, for this set of 55 galaxies, in Tables~\ref{Table1}, \ref{Table2}, and \ref{Table3}, respectively. As one can see from the Tables, for this set of 55 SPARC galaxies, all the values of $\chi^2$ are smaller than 1, indicating a good fit of the theoretical model with the observational data.

With respect to the model parameters, one observes first the very small values of $r_g$ for all galaxies. $r_g$ encodes the baryonic matter contribution coming from the Schwarzschild type component of the metric. On the other hand, the baryonic matter component is also contained in the observational data for $v_{bar}=\sqrt{v_{g}|v_{g}|\!+\!\Upsilon _{d}\!\times
\!v_{d}|v_{d}|\!+\!\Upsilon _{b}\!\times \!v_{b}|v_{b}|\!}$, indicating that the possible contribution of baryonic matter from the Weyl metric is negligibly small.

In the present parameterizations, and investigations  of the Weyl metric we do not assume that the term proportional to $r^2$ is of purely cosmological origin, and hence we do not identify the coefficient $1+C_1/12-r_g/C_2$ with the cosmological constant $\Lambda$. We consider this term as intrinsically belonging to the mathematical structure of the metric, and giving physical effects independent of the cosmological background.

Hence, $C_1$ and $C_2$ are independent quantities, describing the properties of the Weyl vector, and of the scalar field,  which are determined by the local astrophysical properties of the galaxies. Thus, their range and magnitude of values are strictly galaxy-dependent. The constant $C_1$ takes generally negative values, with the exception of a few galaxies. $C_1$ is a dimensionless parameter, taking values in the range $10^7-10^8$. Its mean value for the considered set of bulgeless galaxies is $\left<C_1\right>=4.097\times 10^9$.

\begin{figure*}[htbp]
\centering
\includegraphics[width=0.90\columnwidth]{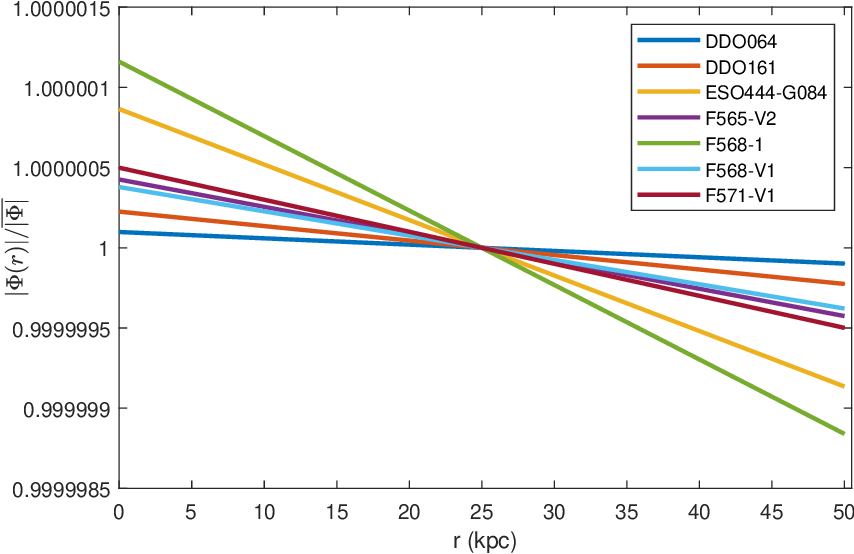} %
\includegraphics[width=0.90\columnwidth]{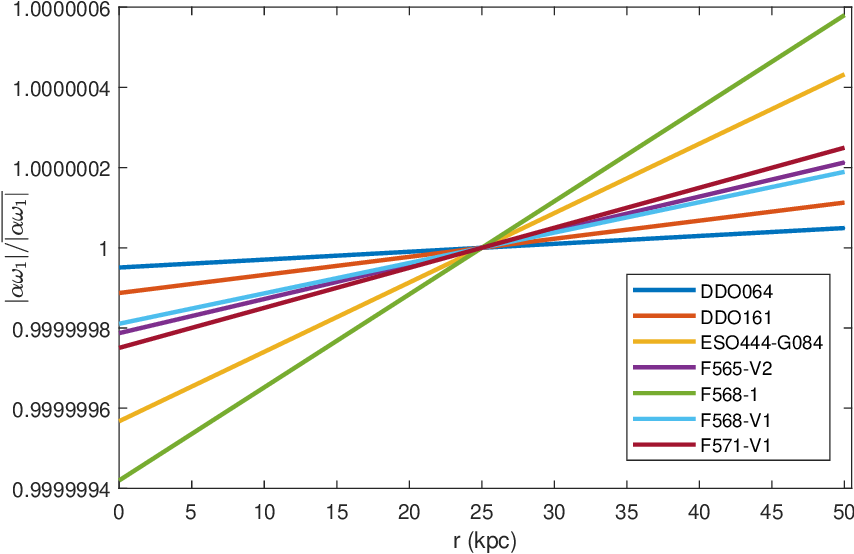} %
\caption{Variation of the scalar field $|\Phi|$ and of $|\alpha~\omega_1|$ inside a selected sample of SPARC galaxies.}
\label{fig:Phi}
\end{figure*}

\begin{figure*}[htbp!]
\centering
\includegraphics[width=0.8\textwidth]{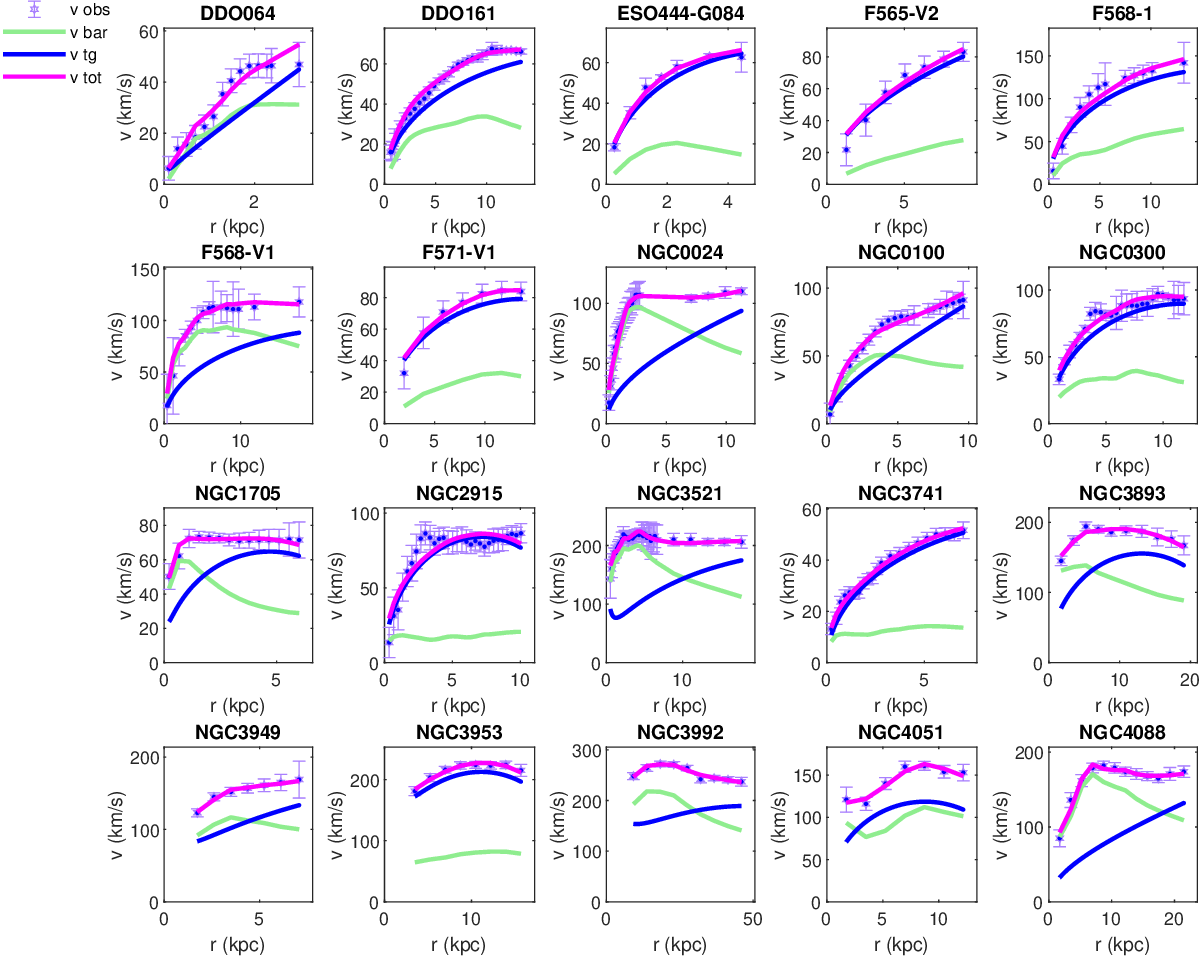}
\caption{Rotation velocities for 20 SPARC galaxies without bulge velocity information data.}
\label{fig_vit1}
\end{figure*}

\begin{figure*}[htbp!]
\centering
\includegraphics[width=0.7\textwidth]{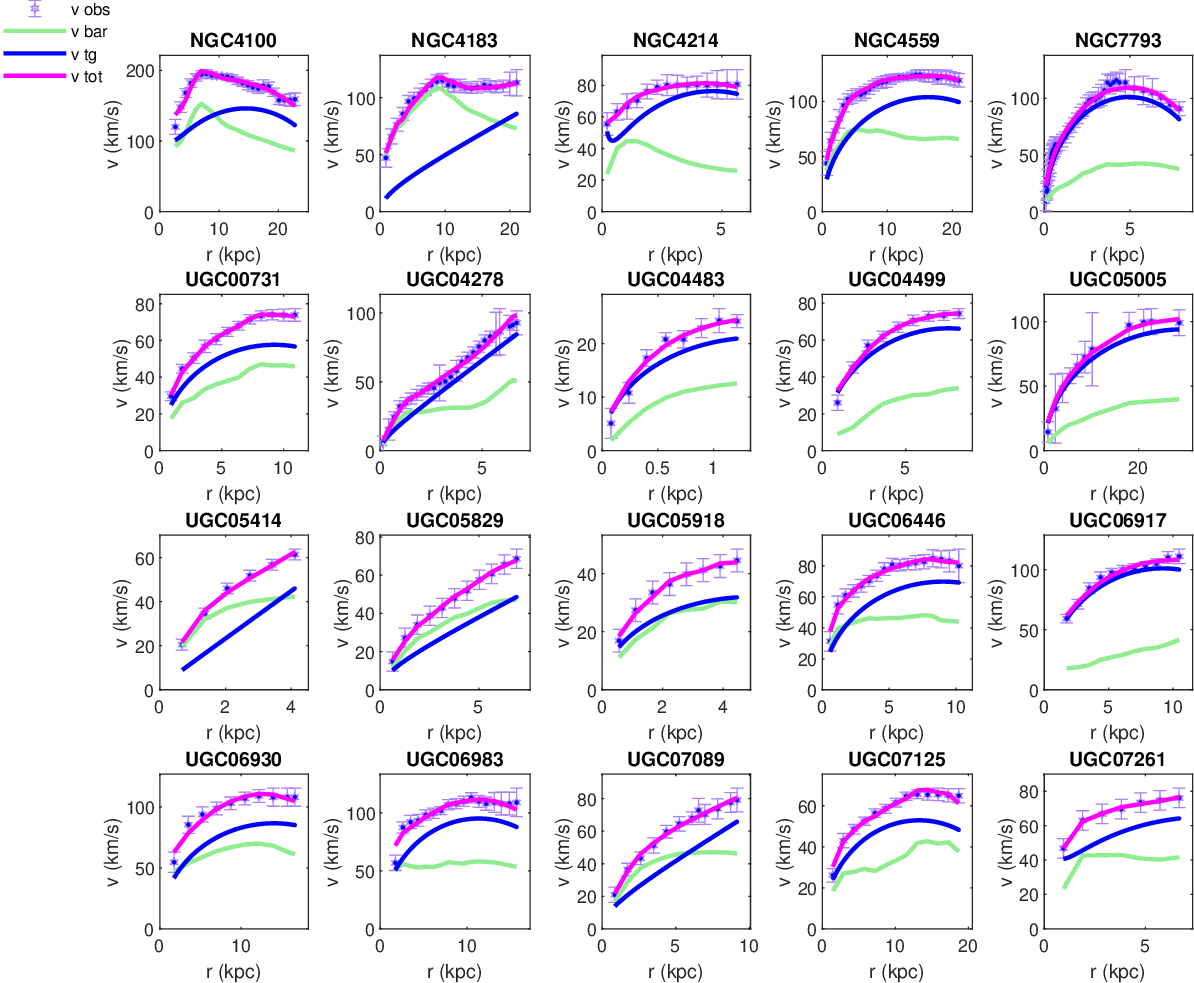}
\caption{Rotation velocities for 20 SPARC galaxies without bulge velocity data.}
\label{fig_vit2}
\end{figure*}

\begin{figure*}[htbp!]
\centering
\includegraphics[width=0.7\textwidth]{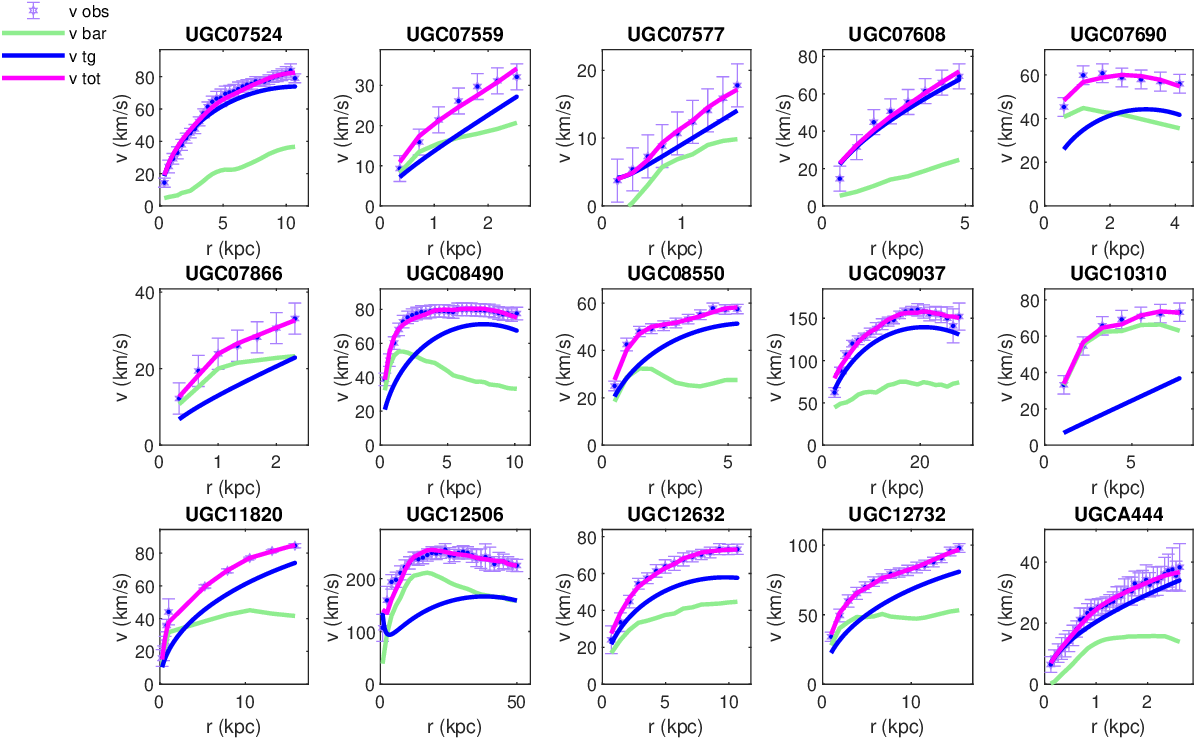}
\caption{Rotation velocities for 15 SPARC galaxies without bulge velocity data. }
\label{fig_vit3}
\end{figure*}

From a physical point of view, as one can see from Eq.~(\ref{40}), $C_1$ describes the distribution of the numerical values of the auxiliary scalar field $\Phi (r)$, which is related, via Eqs.~(\ref{R}) and (\ref{b4}) to the Weyl curvature scalar as $\tilde{R}=\Phi$. This allows to obtain the behavior of $\tilde{R}$ in the dark matter halo of the galaxies, and directly reconstruct the Weyl geometric features of the galactic spacetime. The variations of the scalar field $\Phi$ and of the Weyl vector component $\alpha \omega _1$ inside a small selected sample of galaxies are presented in Fig.~\ref{fig:Phi}.

The constant $C_2$ is a dimensional constant, having dimensions of length. The fitting of the galactic rotation curves gives for $C_2$ values of the same order of magnitude as for $C_1$, $C_2\in \left\{10^7, 10^8\right\}$ kpc, which are much larger values as compared to the extension of the galactic dark matter halo, having values of the order of a few tenths of kiloparsecs. The mean value of the constant $C_2$ is obtained, for the considered set of bulgeless galaxies, as $\left<C_2\right>=3.3478\times 10^8$ kpc.

Since $r<<C_2$, one can estimate the magnitude of scalar field as being given by $\Phi \approx C_1/C_2^2$,  while for the Weyl vector we obtain $\omega _1\approx -2/\alpha C_2$. For the approximate average values of the scalar field we obtain $\left<\Phi\right>=\left<C_1/C_2^2\right>=1.04\times 10^{-8}$ kpc $^{-2}$, and $\left<\Phi\right>=\left<C_1\right>/\left<C_2\right>^2=3.65\times 10^{-8}$ kpc $^{-2}$. In the case of the Weyl vector there is an extra-dependence on the Weyl coupling constant $\alpha$.

Finally, the mass to light ratios $\Upsilon_d$ have relatively small values for all galaxies, exceeding the value one in very few cases. This indicates the realistic nature of the fits, which do not require the introduction of unphysical parameters characterizing baryonic matter.


\subsection{Correlations of the optimal parameters}

In order to check the correlation of the optimal parameters found for each galaxy we made the analysis of the objective function $\chi^2$.
We took  all pairs of two values from the four (respectively five parameters) and we constructed grids of 100X100 points around these optimal values and then we computed the values of $\chi^2$ on these grids keeping the other parameters equal to their optimal values.

\begin{figure*}[htbp!]
\centering
\includegraphics[width=0.47\textwidth]{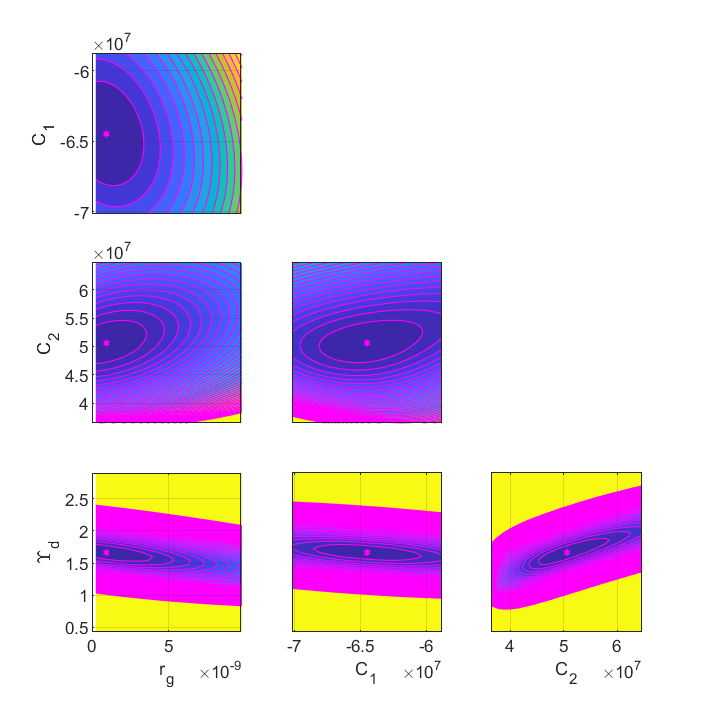}
\includegraphics[width=0.47\textwidth]{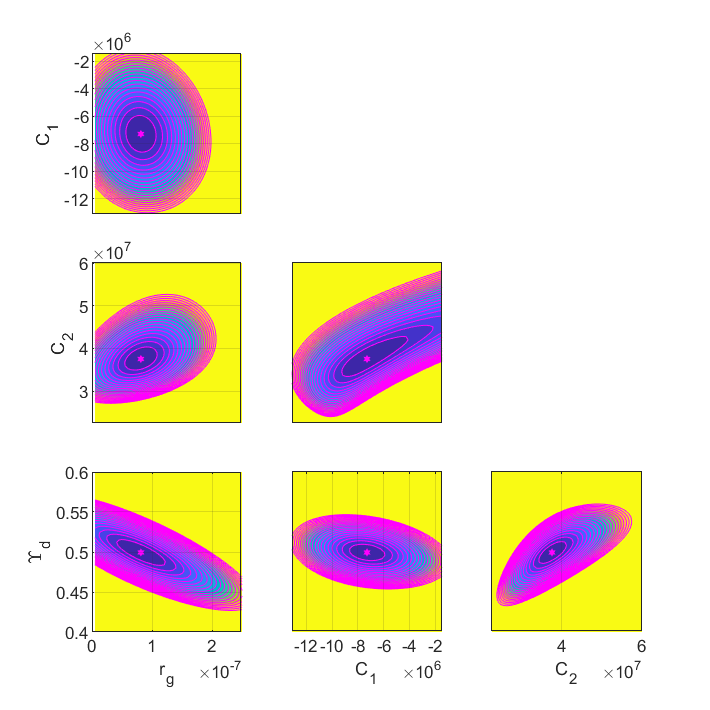}
\caption{Isolines of the $\chi^2$ surfaces around optimal parameters for the galaxy NGC1705 (left panel), and for the galaxy NGC3521 (right panel).}
\label{fig_isolNGC1705}
\end{figure*}

\begin{figure*}[htbp!]
\centering
\includegraphics[width=0.47\textwidth]{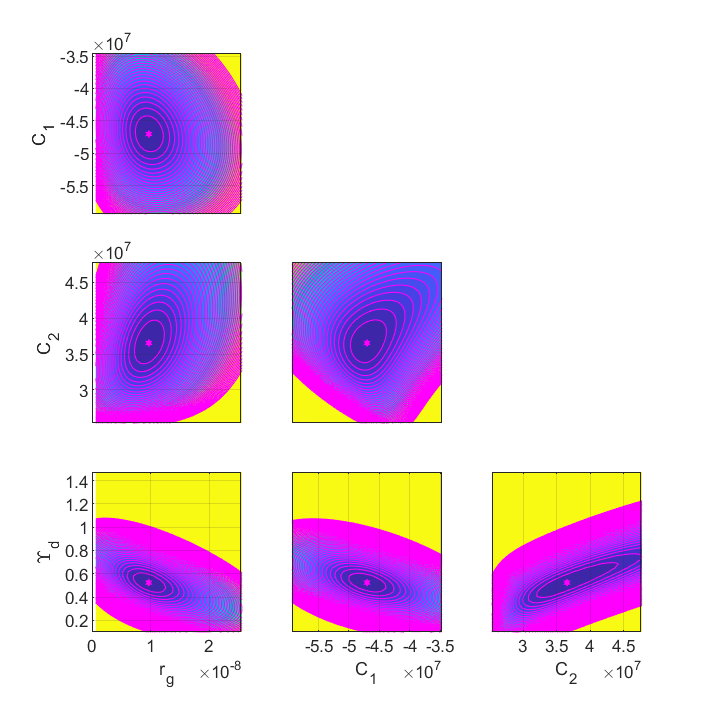}
\includegraphics[width=0.47\textwidth]{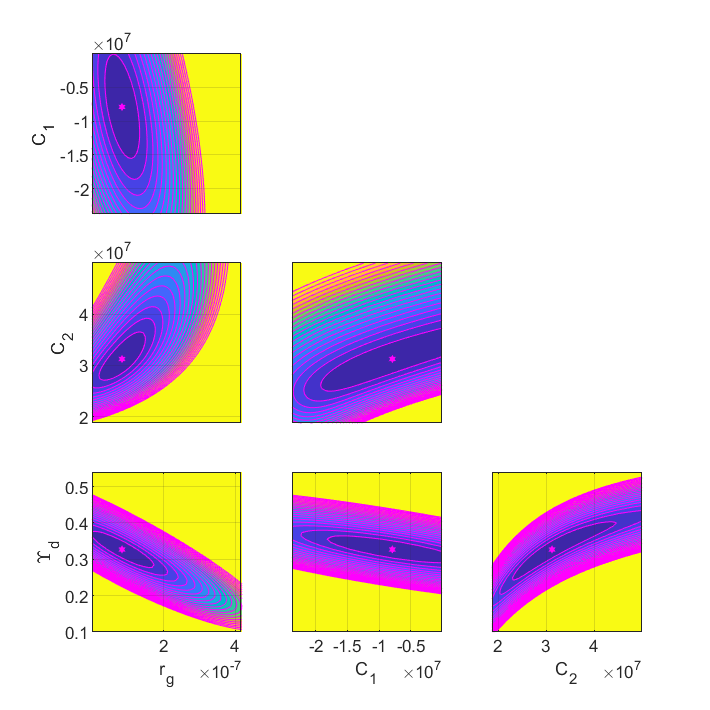}
\caption{ Isolines of the $\chi^2$ surfaces around optimal parameters for the galaxy NGC4214.}
\label{fig_isolNGC4214}
\end{figure*}

We plotted the isolines of the values of $\chi^2$ for four galaxies without bulge velocity data in Figs.~\ref{fig_isolNGC1705},  and \ref{fig_isolNGC4214}, respectively.
Hence, the isolines show the pairs of parameters for which we obtain the same value of $\chi^2$.

We observe that an increase  of $r_g$ corresponds to a decrease of $C_1$ (showing that they are anticorrelated), an increase of $C_2$ (so $r_g$ and $C_2$ are correlated), and  a decrease of $\Upsilon_d$  and $\Upsilon_b$ (so $r_g$ is anti--correlated with $\Upsilon_d$  and $\Upsilon_b$) etc.

Finally, if we increase the value of $\Upsilon_d$ we obtain the same values of $\chi^2$  as for the smaller values
of $\Upsilon_b$ (which is to be expected because if the contribution of the disk to the total velocity increases, the contribution of the bulge must decrease and viceversa).
\color{black}

\begin{table*}
\centering
\begin{tabular}{|l|c|r|r|c|c|}
\hline

Galaxy &  $r_g$ \ \  &  $C_1$ \ \qquad \ \   & $C_2$\ \qquad \ \  & $\Upsilon_d$  & $\chi^2$  \\
\  & ({\rm kpc}) & \   &  ({\rm kpc}) \ \ \ \ \ \ \  &  $ (M_\odot/L_\odot)$   & \  \\
\hline
DDO064 & 3.787e-11 & 5788848096.401 & 508041269.342 & 1.622 & 0.507 \\
\hline
DDO161 & 1.000e-23 & -62562119.155 & 221736484.801 & 0.100 & 0.572 \\
\hline
ESO444-G084 & 1.000e-23 & -61854532.467 & 57806524.365 & 0.100 & 0.754 \\
\hline
F565-V2 & 1.000e-23 & -7088925.839 & 117472239.881 & 0.100 & 0.618 \\
\hline
F568-1 & 1.000e-23 & -14758985.776 & 43093995.313 & 0.770 & 0.988 \\
\hline
F568-V1 & 1.000e-23 & -31757405.870 & 131985481.647 & 3.945 & 0.163 \\
\hline
F571-V1 & 1.000e-23 & -42902459.034 & 100172907.865 & 0.100 & 0.379 \\
\hline
NGC0024 & 1.000e-23 & 26031816.009 & 135988296.772 & 1.865 & 0.348 \\
\hline
NGC0100 & 1.000e-23 & 206024986.877 & 207322173.434 & 0.800 & 0.821 \\
\hline
NGC0300 & 1.000e-23 & -33489958.963 & 65784650.850 & 0.394 & 0.644 \\
\hline
NGC1705 & 9.366e-10 & -64483678.485 & 50613780.152 & 1.664 & 0.121 \\
\hline
NGC2915 & 1.000e-23 & -38060413.192 & 45210549.727 & 0.100 & 0.915 \\
\hline
NGC3521 & 8.146e-08 & -7292886.158 & 37530429.456 & 0.499 & 0.169 \\
\hline
NGC3741 & 1.000e-23 & -86254273.703 & 175001374.119 & 0.462 & 0.366 \\
\hline
NGC3893 & 1.000e-23 & -11155699.476 & 24390676.761 & 0.413 & 0.594 \\
\hline
NGC3949 & 8.391e-08 & -7906447.099 & 31235899.217 & 0.327 & 0.273 \\
\hline
NGC3953 & 5.859e-07 & -6296420.542 & 12002585.314 & 0.100 & 0.396 \\
\hline
NGC3992 & 2.526e-06 & -8063470.748 & 66178422.212 & 0.781 & 0.526 \\
\hline
NGC4051 & 1.000e-23 & -19177151.045 & 28032452.130 & 0.306 & 0.692 \\
\hline
NGC4088 & 1.000e-23 & 32561648.876 & 152931196.546 & 0.402 & 0.494 \\
\hline
\end{tabular}
\caption{Optimal values of the parameters for the galaxies whose rotation velocities are plotted in Fig~\ref{fig_vit1}.}\label{Table1}
\end{table*}


\begin{table*}
\centering
\begin{tabular}{|l|c|r|r|c|c|}
\hline

Galaxy &  $r_g$ \ \  &  $C_1$ \ \qquad \ \   & $C_2$\ \qquad \ \  & $\Upsilon_d$  & $\chi^2$  \\
\  & ({\rm kpc}) & \   &  ({\rm kpc}) \ \ \ \ \ \ \ &  $ (M_\odot/L_\odot)$   & \  \\
\hline
NGC4100 & 2.083e-07 & -13013217.619 & 31954992.408 & 0.534 & 0.650 \\
\hline
NGC4183 & 1.000e-23 & 452475377.873 & 597226941.898 & 1.637 & 0.246 \\
\hline
NGC4214 & 9.673e-09 & -47027873.229 & 36541360.348 & 0.524 & 0.076 \\
\hline
NGC4559 & 1.000e-23 & -25031486.089 & 67875858.531 & 0.385 & 0.098 \\
\hline
NGC7793 & 1.000e-23 & -26348417.531 & 21709425.077 & 0.100 & 0.576 \\
\hline
UGC00731 & 1.000e-23 & -81127914.722 & 125161386.432 & 5.000 & 0.101 \\
\hline
UGC04278 & 1.000e-23 & 2200684669.177 & 366439884.486 & 0.923 & 0.454 \\
\hline
UGC04483 & 1.000e-23 & -603007109.979 & 141169326.943 & 0.100 & 0.771 \\
\hline
UGC04499 & 1.000e-23 & -61295667.386 & 77406914.217 & 0.100 & 0.699 \\
\hline
UGC05005 & 1.000e-23 & -30551927.021 & 149493907.016 & 0.100 & 0.177 \\
\hline
UGC05414 & 1.000e-23 & 80431750353.603 & 2280981528.847 & 1.061 & 0.778 \\
\hline
UGC05829 & 1.000e-23 & 1349304293.380 & 602130434.197 & 2.522 & 0.053 \\
\hline
UGC05918 & 1.000e-23 & -262450081.090 & 220573114.601 & 2.327 & 0.128 \\
\hline
UGC06446 & 1.000e-23 & -55187096.744 & 82990617.223 & 1.811 & 0.195 \\
\hline
UGC06917 & 1.000e-23 & -26428443.948 & 40479816.606 & 0.100 & 0.535 \\
\hline
UGC06930 & 1.000e-23 & -35940098.272 & 84004991.006 & 0.799 & 0.501 \\
\hline
UGC06983 & 1.000e-23 & -29767444.028 & 56137365.807 & 0.782 & 0.629 \\
\hline
UGC07089 & 1.000e-23 & 851698942.747 & 458950441.378 & 0.697 & 0.157 \\
\hline
UGC07125 & 1.000e-23 & -96323598.406 & 212796623.635 & 0.340 & 0.601 \\
\hline
UGC07261 & 1.564e-08 & -65132116.799 & 87859036.829 & 0.688 & 0.028 \\
\hline
\end{tabular}
\caption{ Optimal values of the parameters for the galaxies whose rotation velocities are plotted in Fig~\ref{fig_vit2}.}\label{Table2}
\end{table*}


\begin{table*}
\centering
\begin{tabular}{|l|c|r|r|c|c|}
\hline
Galaxy &  $r_g$ \ \  &  $C_1$ \ \qquad \ \   & $C_2$\ \qquad \ \  & $\Upsilon_d$  & $\chi^2$  \\
\  & ({\rm kpc}) & \   & ({\rm kpc}) \ \ \ \ \ \ \  & \  $ (M_\odot/L_\odot)$   & \  \\
\hline
UGC07524 & 1.000e-23 & -49350136.845 & 90321980.010 & 0.100 & 0.517 \\
\hline
UGC07559 & 1.000e-23 & 5370120200.786 & 764005723.368 & 0.658 & 0.515 \\
\hline
UGC07577 & 4.381e-11 & 66656987997.450 & 3104152296.378 & 0.100 & 0.030 \\
\hline
UGC07608 & 1.000e-23 & 51193449.496 & 111393221.187 & 0.100 & 0.581 \\
\hline
UGC07690 & 1.000e-23 & -137870331.963 & 71148385.044 & 0.784 & 0.506 \\
\hline
UGC07866 & 1.000e-23 & 3077280935.014 & 724351653.383 & 1.258 & 0.047 \\
\hline
UGC08490 & 1.000e-23 & -53025454.948 & 67806785.219 & 1.350 & 0.212 \\
\hline
UGC08550 & 1.000e-23 & -101438913.618 & 99962034.196 & 1.126 & 0.508 \\
\hline
UGC09037 & 1.000e-23 & -13869977.750 & 48003507.625 & 0.100 & 0.939 \\
\hline
UGC10310 & 1.000e-23 & 56281824440.461 & 4578269948.117 & 2.396 & 0.189 \\
\hline
UGC11820 & 1.000e-23 & -28209554.429 & 214936509.292 & 1.663 & 0.460 \\
\hline
UGC12506 & 3.149e-07 & -9874154.696 & 63305217.998 & 1.370 & 0.554 \\
\hline
UGC12632 & 1.000e-23 & -80421187.018 & 130076378.298 & 1.413 & 0.220 \\
\hline
UGC12732 & 1.000e-23 & -27533667.612 & 166645457.180 & 1.827 & 0.131 \\
\hline
UGCA444 & 1.000e-23 & 103957781.561 & 223885591.848 & 0.100 & 0.056 \\
\hline
\end{tabular}
\caption{ Optimal values of the parameters for the galaxies whose rotation velocities are plotted in Fig~\ref{fig_vit3}.}\label{Table3}
\end{table*}


\subsection{Fitting of the galaxies with bulge velocity data}

\begin{figure*}[htbp!]
\centering
\includegraphics[width=0.6\textwidth]{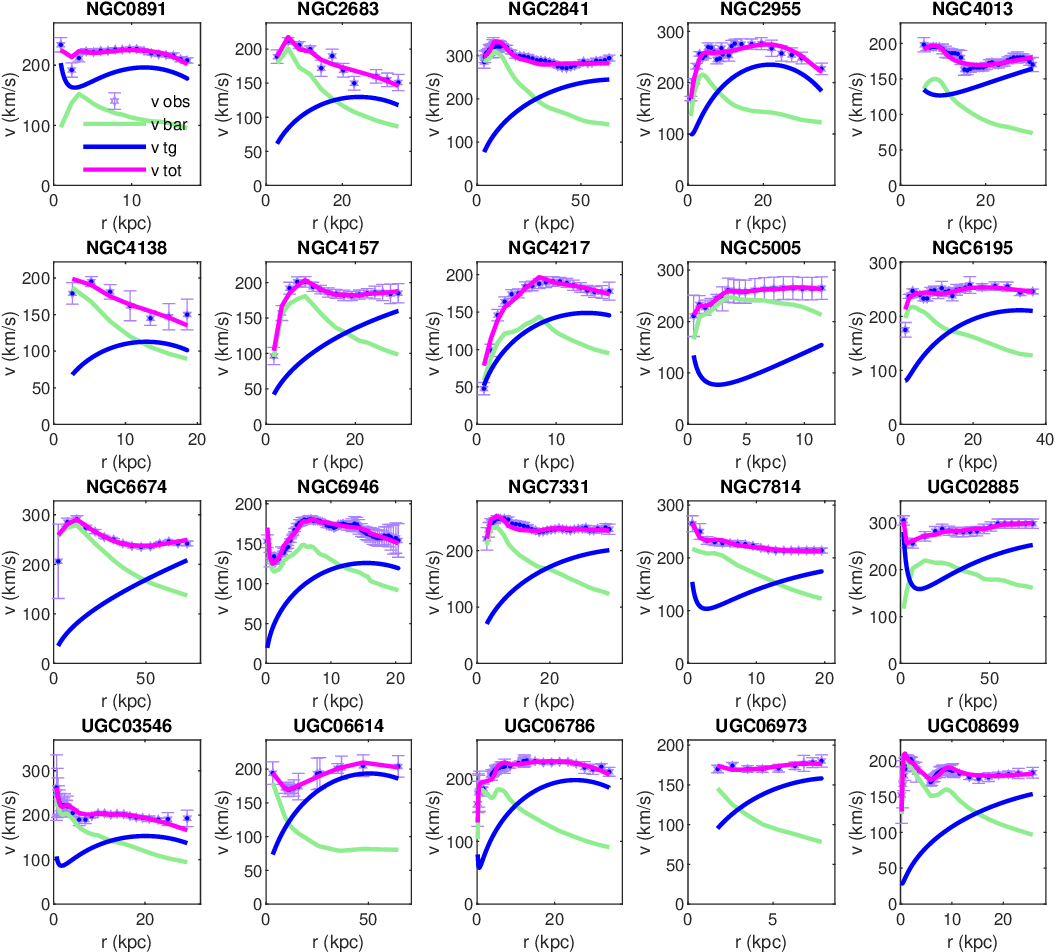}
\caption{Rotation velocities for 20 galaxies with bulge velocity data.}
\label{fig_vitbb}
\end{figure*}

The results of the fitting of a sample of 20 galaxies with bulge velocities are presented in Fig.~\ref{fig_vitbb}. The optimal values of the model parameters, and of the mass to light ratios are presented, together with the $\chi^2$ values, in Table~\ref{table_param_b}. Similarly to the bulgeless case, the parameter $r_g$ takes very small values, indicating that the baryonic matter component, as obtained from the corresponding baryonic velocity distribution, is enough for the interpretation of the data, and no extra baryonic component is needed.

The numerical values of the constants $C_1$ and $C_2$ are in the same quantitative range as in the bulgeless case. The constant $C_1$ takes again negative values, while $C_2$ is always positive, and takes large values, exceeding those taken by the radial coordinate inside the galactic halo. The mean value of the constant $C_1$ for the galxies with bulge velocity contribution is $\left<C_1\right>=2.01\times 10^7$, while the mean value of the constant $C_2$ is $\left<C_2\right>=5.79\times 10^7$ kpc$^{-2}$. For the mean values of the scalar field we obtain $\left<\Phi\right>=\left<C_1/C_2^2\right>=9.08\times 10^{-9}$ kpc $^{-2}$, and $\left<\Phi\right>=\left<C_1\right>/\left<C_2\right>^2=5.99\times 10^{-9}$ kpc $^{-2}$, respectively.

The two mass to light ratios have acceptable values, and only in a few cases values of $\Upsilon _d$ bigger than one are necessary for the fitting. The distribution of the $\chi^2$ function indicates a values ranging from $\chi^2=0.022$ to $\chi^2=2.046$, with most of values in the range $\chi ^2<1$. These results show that the Weyl geometric gravity model can provide acceptable fits for the observational data even for this set of observations.

\begin{figure*}[htbp!]
\noindent
\includegraphics[width=0.47\textwidth]{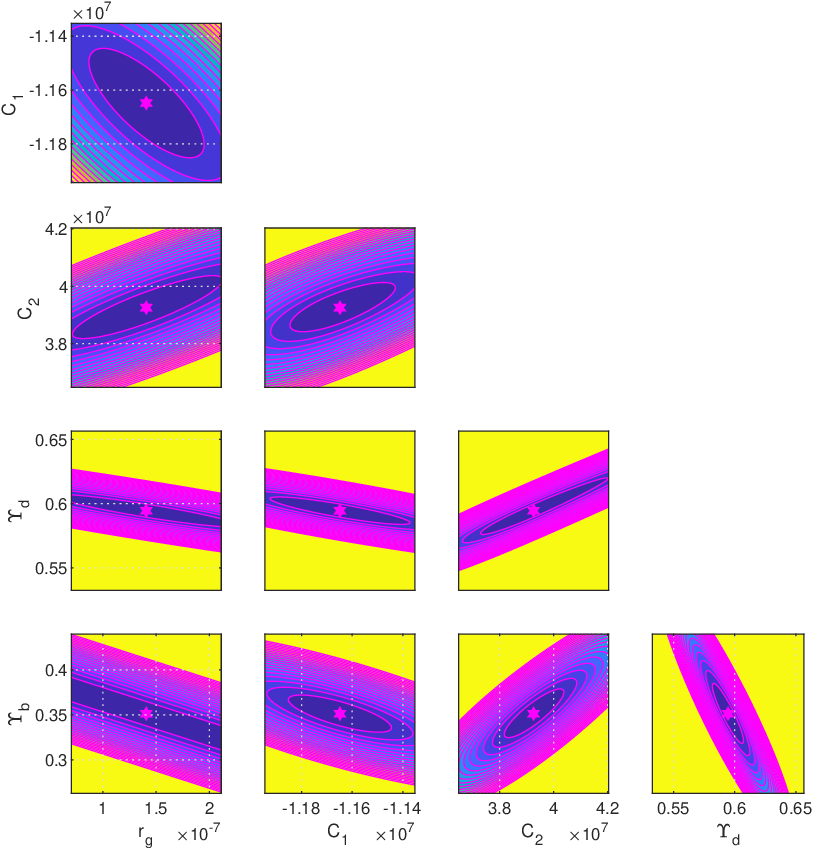}
\hspace{0.1cm}
\includegraphics[width=0.47\textwidth]{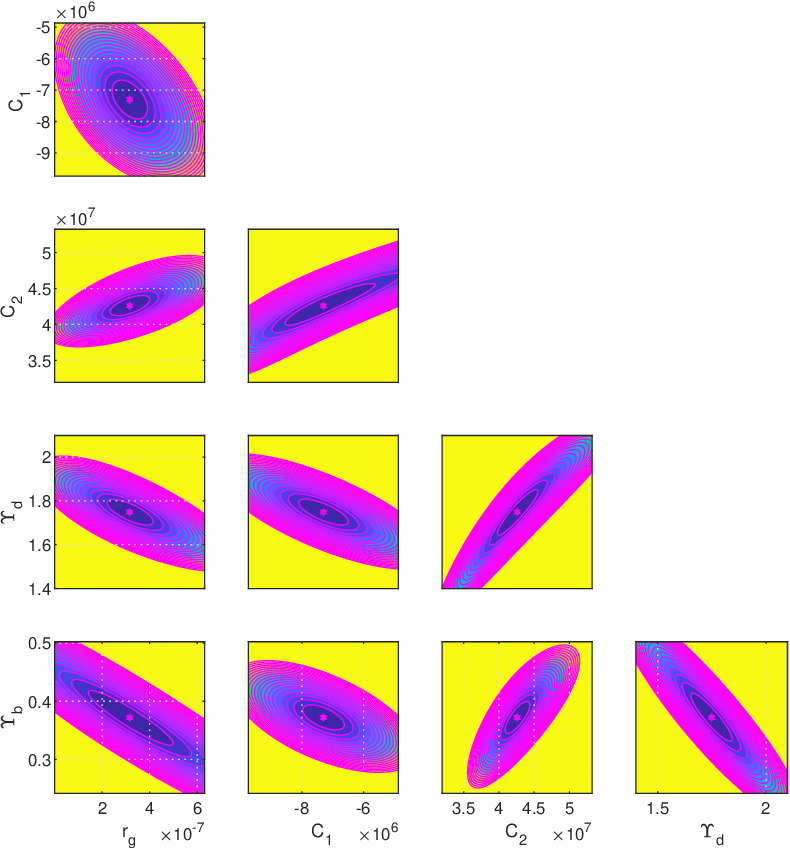}

\caption{Isolines of the $\chi^2$ surfaces around optimal parameters for the galaxy UGC03546 (left panel) and for the galaxy NGC7814 (right panel).}
\label{fig_isolUGC03546}
\end{figure*}

\begin{figure*}[htbp!]
\noindent
\includegraphics[width=0.47\textwidth]{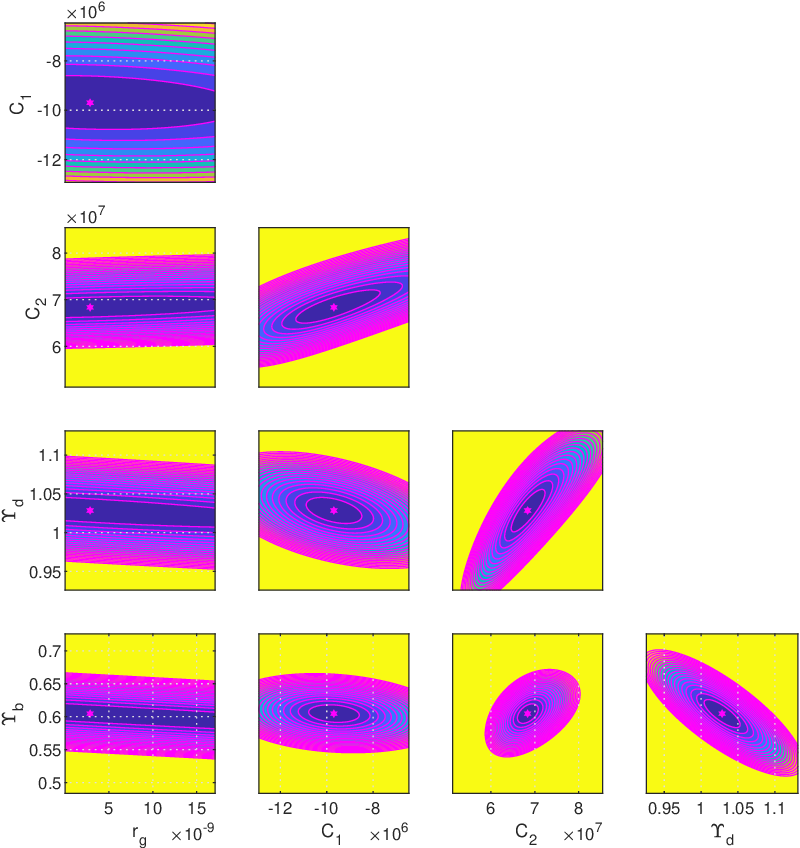}
\hspace{0.1cm}
\includegraphics[width=0.47\textwidth]{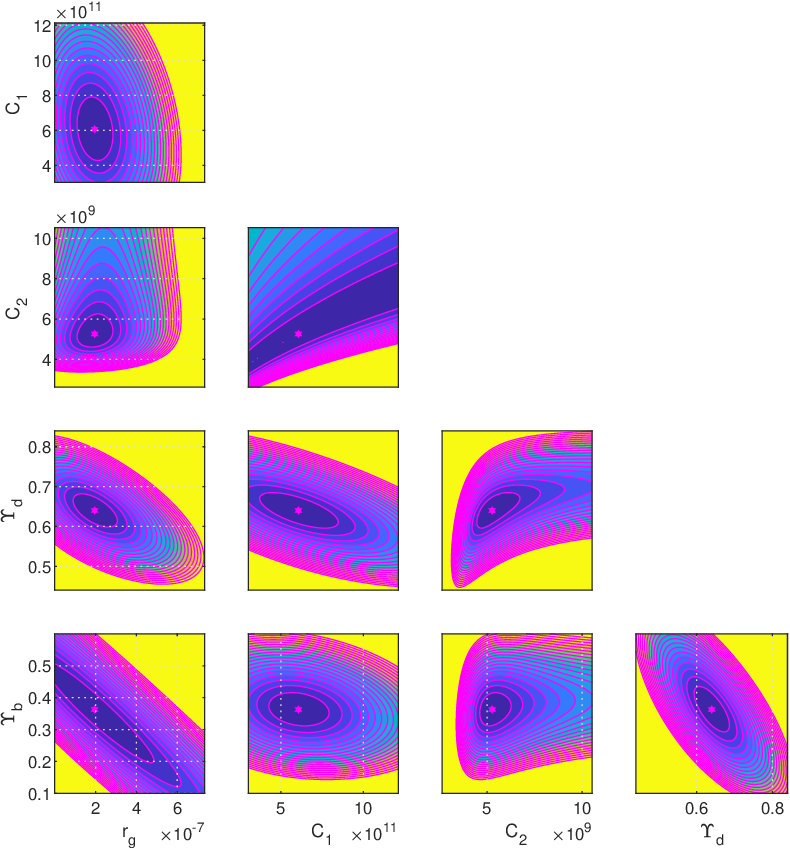}
\caption{Isolines of the $\chi^2$ surfaces around optimal parameters for the galaxy UGC08699 and for the galaxy NGC5005.}
\label{fig_isolUGC08699}
\end{figure*}

The isolines of the $\chi ^2$ distribution for a small set of galaxies with bulge velocity are presented in Figs.~\ref{fig_isolUGC03546},  and \ref{fig_isolUGC08699}, respectively.

\begin{table*}
\centering
\begin{tabular}{|l|c|r|r|c|c|c|}
\hline
Galaxy &  $r_g$ \ \  &  $C_1$ \ \qquad \ \   & $C_2$\ \qquad \ \  & $\Upsilon_d$  &$\Upsilon_b$ & $\chi^2$  \\
\  & ({\rm kpc}) & \   & ({\rm kpc}) \ \ \ \ \ \ \ & $ (M_\odot/L_\odot)$   & $ (M_\odot/L_\odot)$ &  \\
\hline
NGC0891 & 7.406e-07 & -7548219.795 & 15253073.451 & 0.197 & 0.100 & 1.234 \\
\hline
NGC2683 & 1.000e-23 & -16000065.073 & 65051713.809 & 0.753 & 0.100 & 1.604 \\
\hline
NGC2841 & 1.080e-13 & -4496210.516 & 50623362.039 & 1.246 & 0.884 & 1.748 \\
\hline
NGC2955 & 1.102e-07 & -4902267.287 & 17830128.465 & 0.100 & 0.674 & 1.687 \\
\hline
NGC4013 & 1.616e-06 & -5131504.307 & 97579798.795 & 0.494 & 0.100 & 0.743 \\
\hline
NGC4138 & 1.000e-23 & -21176828.022 & 45559450.954 & 0.855 & 0.100 & 2.046 \\
\hline
NGC4157 & 1.001e-23 & -5894284.347 & 86512981.587 & 0.503 & 0.100 & 0.265 \\
\hline
NGC4217 & 1.000e-23 & -12151305.398 & 28073861.051 & 1.181 & 0.123 & 1.854 \\
\hline
NGC5005 & 1.913e-07 & 232753862.401 & 124443725.256 & 0.614 & 0.365 & 0.022 \\
\hline
NGC6195 & 8.931e-08 & -6061759.524 & 33240834.431 & 0.100 & 0.687 & 1.647 \\
\hline
NGC6674 & 4.387e-11 & 5395586.701 & 177645686.226 & 1.248 & 1.598 & 1.341 \\
\hline
NGC6946 & 1.000e-23 & -17033560.155 & 43947101.122 & 0.475 & 0.519 & 1.507 \\
\hline
NGC7331 & 1.001e-23 & -6500574.938 & 47160989.502 & 0.421 & 0.100 & 0.845 \\
\hline
NGC7814 & 3.157e-07 & -7302964.454 & 42596044.256 & 1.749 & 0.372 & 0.408 \\
\hline
UGC02885 & 2.890e-06 & -3936362.944 & 70070401.820 & 0.778 & 0.100 & 0.173 \\
\hline
UGC03546 & 1.407e-07 & -11648346.858 & 39256891.791 & 0.595 & 0.351 & 0.888 \\
\hline
UGC06614 & 5.654e-08 & -7209010.216 & 60761899.115 & 0.100 & 0.521 & 0.023 \\
\hline
UGC06786 & 2.048e-08 & -6897154.187 & 29136907.700 & 0.766 & 0.752 & 1.609 \\
\hline
UGC06973 & 1.001e-23 & -10669113.237 & 15495630.244 & 0.177 & 0.100 & 0.751 \\
\hline
UGC08699 & 2.852e-09 & -9693947.277 & 68352653.068 & 1.029 & 0.605 & 0.665 \\
\hline
\end{tabular}
\caption{Optimal values of the parameters for the galaxies with bulge whose rotation velocities are plotted in Fig~\ref{fig_vitbb}.}
\label{table_param_b}
\end{table*}

\subsection{Optimal parameters distribution}

In the following we consider the distributions of the various statistical parameters of the Weyl geometric gravity dark matter model.

\begin{figure}[htbp!]
\centering
\includegraphics[width=0.55\textwidth]{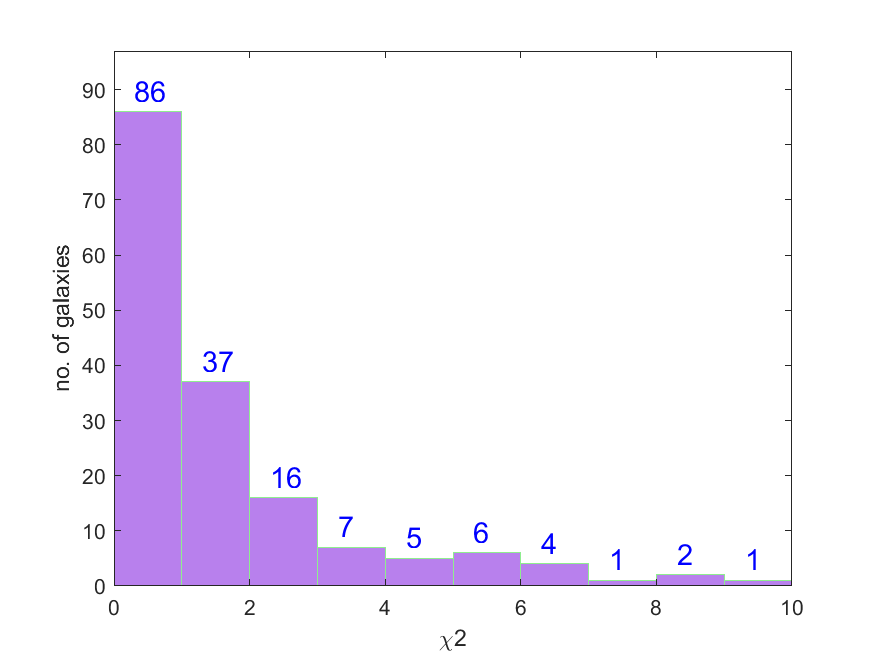}
\caption{Histogram of the objective function $\chi^2$ values (for $\chi^2 \le 10$).}
\label{fig_chi2}
\end{figure}

\paragraph{$\chi^2$ distribution of the SPARC sample.} Fig.~\ref{fig_chi2} presents the distribution of the values of the objective function $\chi^2$ for the total number of the considered galaxies for which $\chi^2 \le 10$.  One can observe that there are 86 galaxies with $\chi^2 <1$, 37 galaxies with $\chi^2$ between 1 and 2 and 16 galaxies with $\chi^2$ between 2 and 3, which means a total of 139 galaxies for which the value of $\chi^2$ is less than 3 (from the 171 galaxies that we have analyzed, {\it i.e.}, representing a percent of  more than 80\%). From this point of view we can consider that the Weyl geometric gravity model of the dark matter gives an acceptable description of the observational data of the galactic rotation curves.

\begin{figure*}[h]
\begin{center}
\includegraphics[scale=0.55]{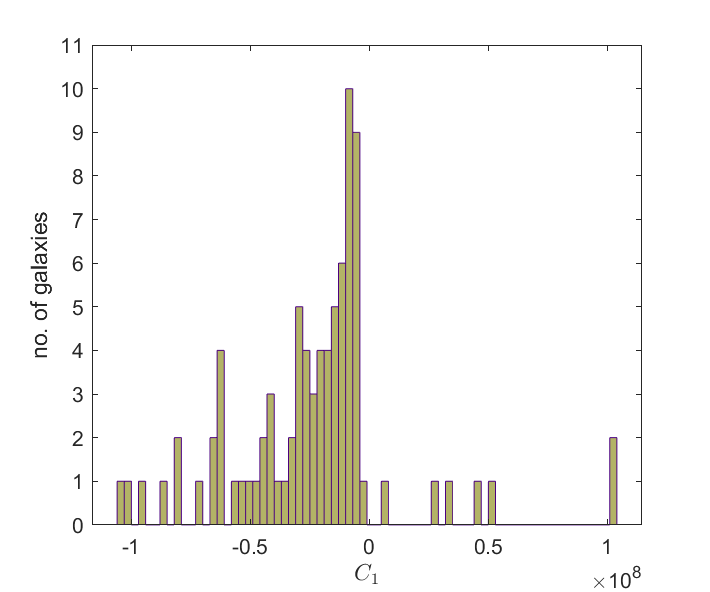}
\includegraphics[scale=0.55]{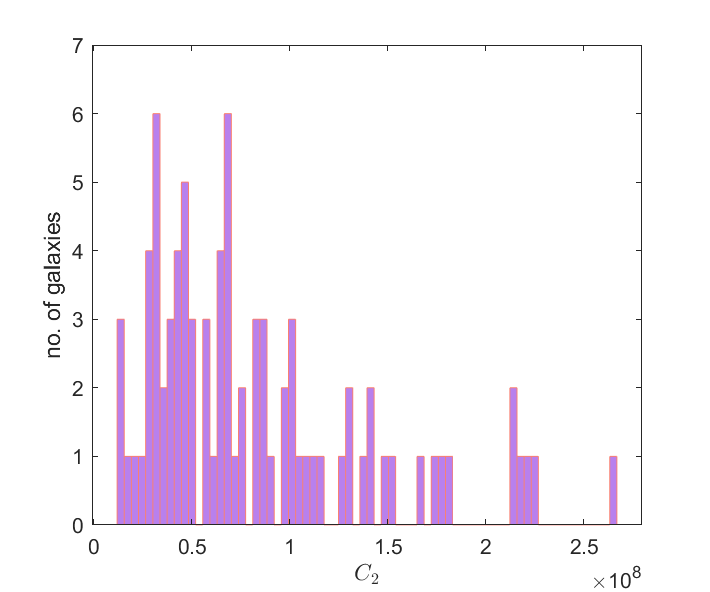}
\caption{Histogram of the optimal values of the parameters $C_1$ and $C_2$.}
\end{center}
\label{figC1andC2}
\end{figure*}

\paragraph{Distribution of the optimal values of $C_1$ and $C_2$} In Fig.~\ref{figC1andC2} the distributions of the optimal values of $C_1$ and $C_2$ are presented.
For better visibility we have omitted few values larger than $1.2 \times 10^8$ for $C_1$, respectively larger than $3 \times 10^8$ for $C_2$.

\begin{figure*}[h]
\begin{center}
\includegraphics[scale=0.55]{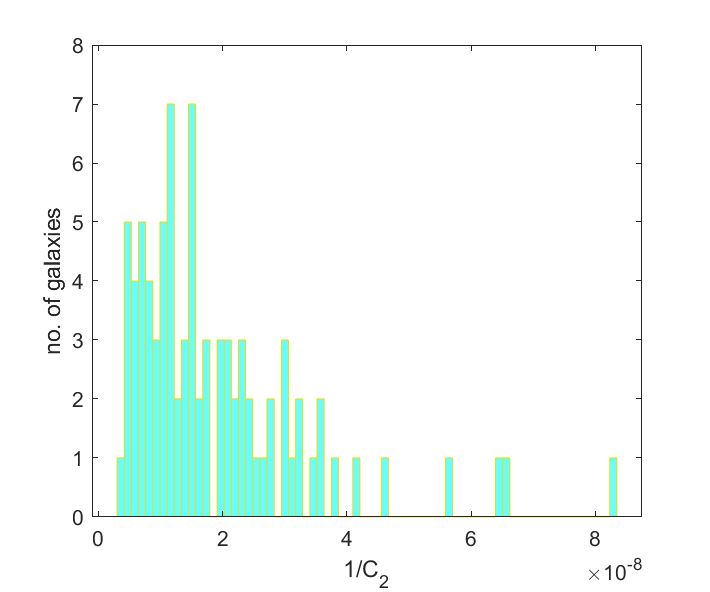}
\includegraphics[scale=0.55]{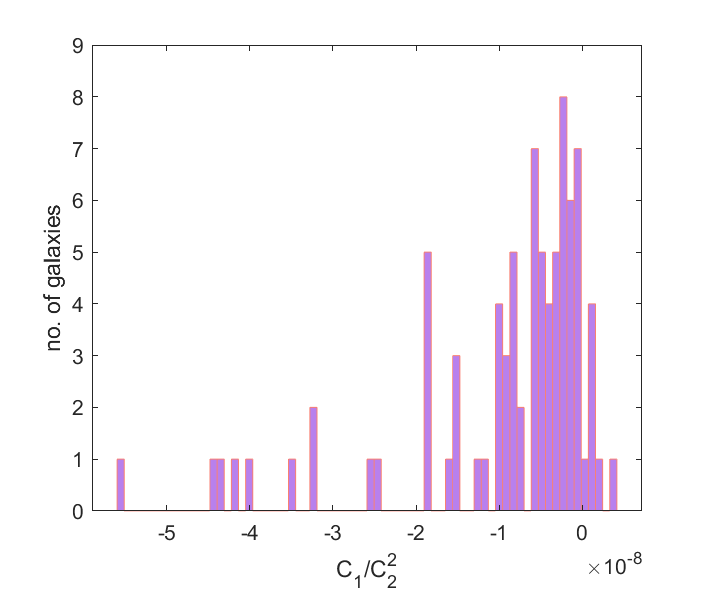}
\caption{Histogram of the optimal values of the quantities $1/C_2$ and $C1/C_2^2$.}
\label{fig_1peC2andC1peC2la2}
\end{center}
\end{figure*}

\paragraph{Distribution of the optimal values of $1/C_2$ and $C_1/C_2^2$} In Fig.~ \ref{fig_1peC2andC1peC2la2} the distributions of the optimal values of  $1/C_2$ and $C_1/C_2^2$ are presented.

\subsection{Correlations between the parameters of the model, and the astrophysical quantities}

We proceed now to the investigation of the problem of the existence of possible correlations between the parameters of the Weyl geometric dark matter model, and between these parameters and the astrophysical quantities describing galactic properties.

\begin{figure*}[h]
\begin{center}
\includegraphics[scale=0.55]{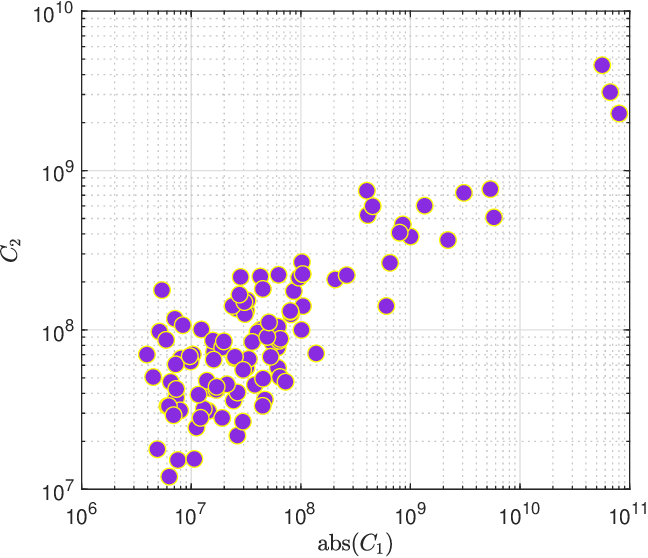}
\hspace{1cm}
\includegraphics[scale=0.55]{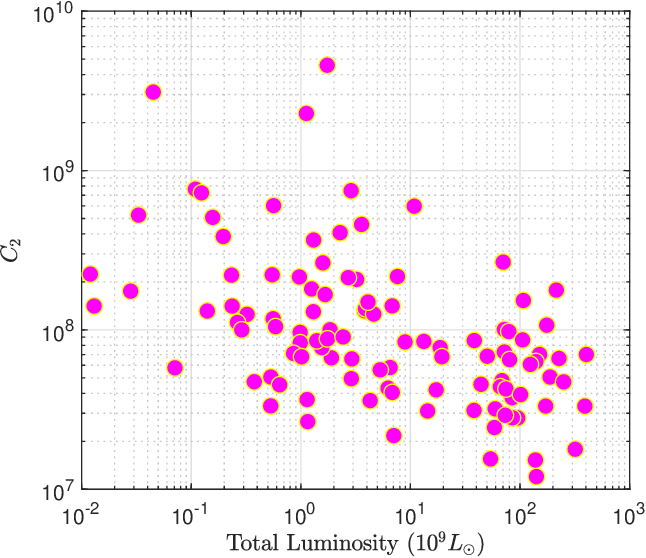}
\caption{The optimal values of $C_2$ versus $abs(C_1)$ (left panel) and $C_2$ versus total galactic Luminosity (right panel). The correlation
coefficients are  0.8817, and -0.1618, respectively.}
\label{C1C2andTLC2}
\end{center}
\end{figure*}

\paragraph{Correlation between $C_1$ and $C_2$} The first plot of Fig.~\ref{C1C2andTLC2} presents the  optimal values of $C_2$ versus the absolute values of $C_1$, these values being correlated with the Pearson correlation coefficient 0.8817. The second plot shows the  optimal values of $C_2$ versus the total luminosity of the corresponding galaxies. From the Figure one can see that they are slightly anti-correlated.

\begin{figure*}[h]
\begin{center}
\hspace{-1cm}
\includegraphics[scale=0.55]{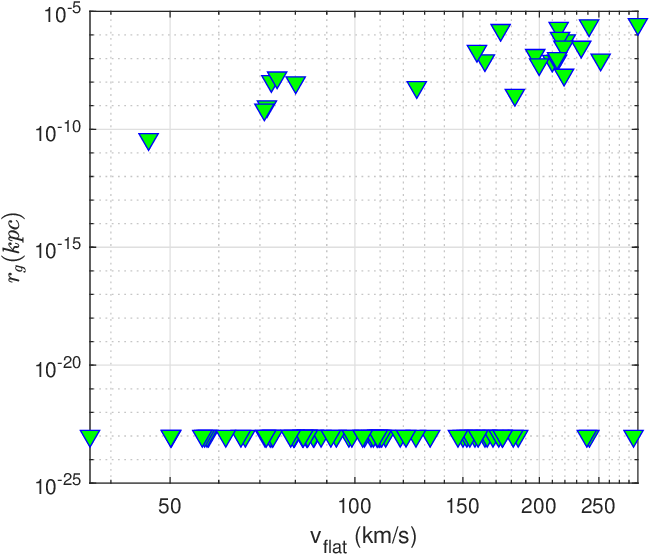}
\hspace{1cm}
\includegraphics[scale=0.55]{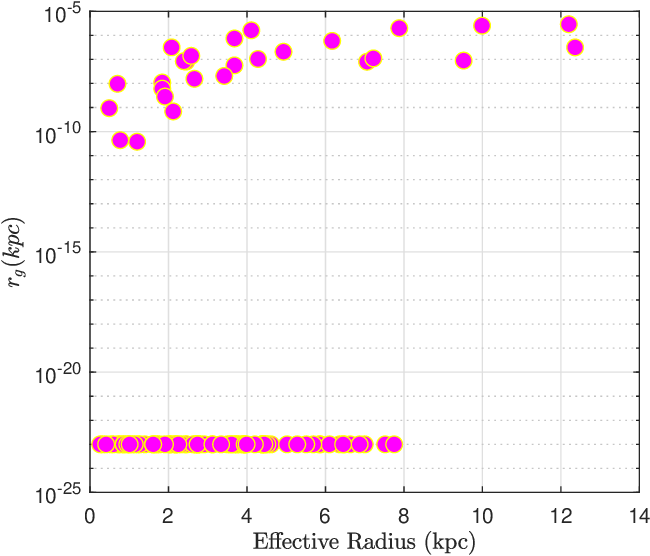}
\caption{ The optimal values of $r_g$ versus $v_{flat}$ (left panel), respectively  $r_g$ versus the effective radius (right panel), the correlation
coefficients being  0.4584 and 0.4875.}
\label{rg_vflat_and_rg_effr}
\end{center}
\end{figure*}

\paragraph{Correlation of $r_g$ with $v_{flat}$ and the effective radius} Fig.~\ref{rg_vflat_and_rg_effr} shows a log-log plot of the optimal values of $r_g$ versus the asymptotic value of velocity $v_{flat}$ that indicate their medium correlation (the correlation
coefficient being  0.4584). The right panel of Fig.~~\ref{rg_vflat_and_rg_effr} presents a semi-logarithmic  plot of $r_g$ versus the effective radius of the corresponding galaxies (again they are medium correlated with
the correlation coefficient 0.4875).

\begin{figure*}[h]
\begin{center}
\hspace{-1cm}
\includegraphics[scale=0.55]{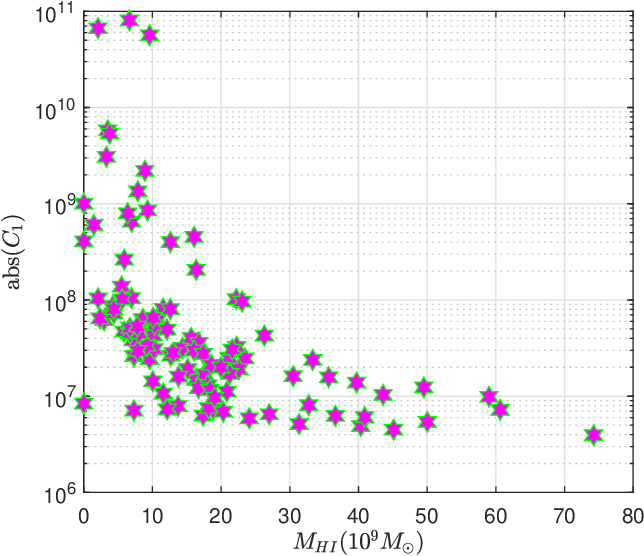}
\hspace{1cm}
\includegraphics[scale=0.55]{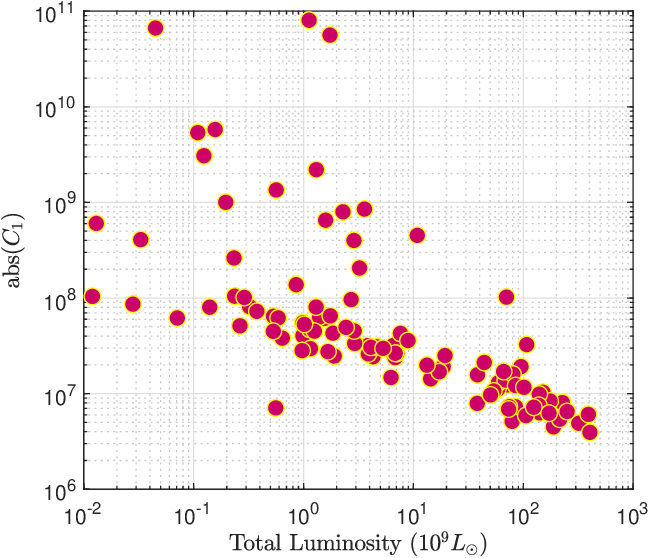}
\caption{The optimal values of $abs(C_1)$ versus $M_{HI}$ (left panel), respectively $abs(C_1)$ versus total galactic Luminosity (right panel).}
\label{absC1_THImass_and_absC1_TL}
\end{center}
\end{figure*}

\paragraph{Correlating $C_1$ with the hydrogen mass and galactic luminosity} In Fig.~\ref{absC1_THImass_and_absC1_TL} we have represented the absolute value of the optimal values of $C_1$ versus the total mass of hydrogen $M_{HI}$, respective versus the total luminosities of the galaxies. One can see that these quantities are slightly anticorrelated.

\begin{figure*}[h]
\begin{center}
\hspace{-1cm}
\includegraphics[scale=0.55]{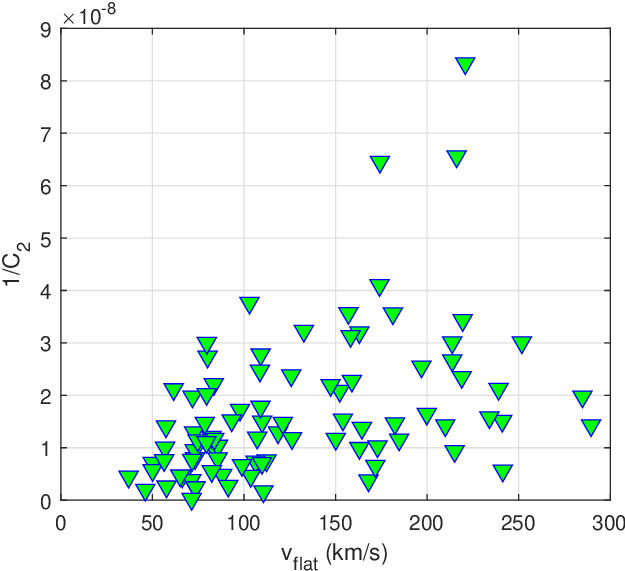}
\hspace{1cm}
\includegraphics[scale=0.55]{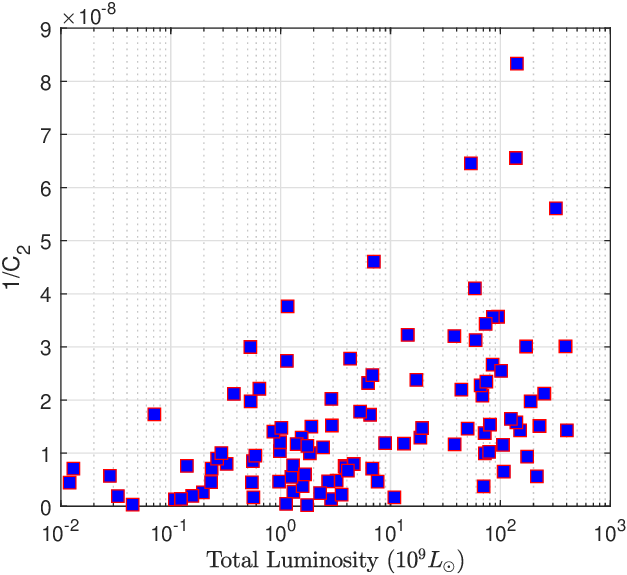}
\caption{The optimal values of $1/C_2$ versus $v_{flat}$ (left panel), respectively $1/C_2$ versus total galactic Luminosity (right panel), the correlation
coefficients being  0.432, respectively 0.3645.}
\label{1peC2_vflat}
\end{center}
\end{figure*}

\paragraph{Correlation of $1/C_2$ with $v_{flat}$ and galactic luminosity} The left panel of Fig.~\ref{1peC2_vflat} shows a plot of the optimal values of $1/C_2$ versus the asymptotic value of velocity $v_{flat}$ that indicate their medium correlation (the correlation
coefficient being  0.432). The right panel represents the optimal values $1/C_2$ versus the total luminosities of the galaxies (semilog plot). One can see that these quantities are slightly correlated their correlation coefficient being 0.3645.

\begin{figure}[h]
\begin{center}
\hspace{-1cm}
\includegraphics[scale=0.57]{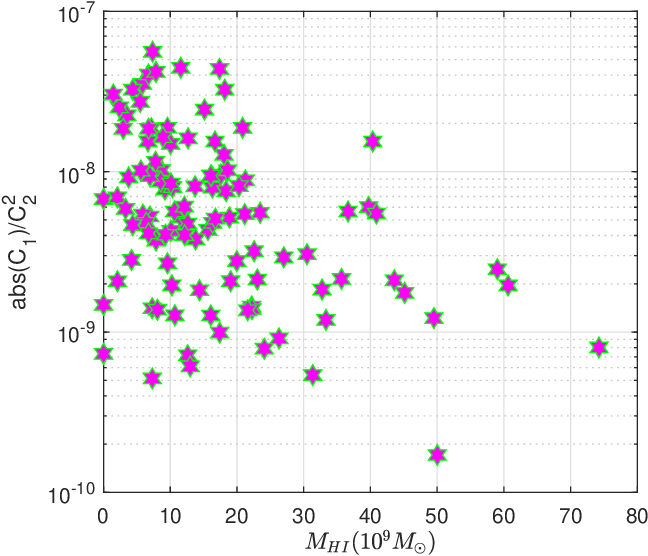}
\caption{The optimal values of $abs(C1)/C_2^2$ versus $M_{HI}$.}
\label{THImass_absC1peC2la2}
\end{center}
\end{figure}

\paragraph{Correlating $abs(C_1)/C_2^2$ with the hydrogen mass} In Fig.~\ref{THImass_absC1peC2la2} we have represented the optimal quantities $abs(C_1)/C_2^2$ versus the total mass of hydrogen $M_{HI}$. One can see that these quantities are slightly anticorrelated the correlation coefficients being -0.3036.

\color{black}

\section{Discussions and final remarks}\label{sect4}

Weyl geometry has experienced recently a strong revival, mostly due to its beautiful mathematical structure, and attractive physical ideas that could be implemented by using its formalism. A significant step in this direction was undertaken by Dirac \cite{D1,D2}, who tried to reformulate Weyl's theory from a physical point of view, by introducing a real scalar field $\beta$ of weight $w(\phi)=-1$. The action proposed by Dirac is conformally invariant, and in the cosmological model based on the action introduced in \cite{D2}, the presence of the Dirac scalar gauge field leads to the creation of matter at the very beginning of the Universe. Moreover, in the late expansionary stages of the Universe, the Dirac scalar gauge field may represent  the dark energy that triggers present accelerated cosmic expansion of the Universe. Dirac's generalization of Weyl's theory is one of the first modifications in which a new scalar degree of freedom was added to the theory in order to extend the initial vector-tensor formulation.

The idea of conformal invariance, central in Weyl geometry, is extensively used in the Conformal Cyclic Cosmology (CCC) model \cite{P1,P2}, in which the basic assumption is that the Universe exists as o set of eons. Eons are geometric structures representing time oriented spacetimes. Eons possess spacelike null infinities as a result of their conformal compactification. The general physical and cosmological implications of the CCC model were studied in \cite{P3,P4,P5}.

In \cite{Ho6} Gerard 't Hooft suggested that conformal symmetry is an exact symmetry that is spontaneously broken during the evolution of the Universe. Hence conformal symmetry  could be of equal importance to the Lorentz invariance of natural laws. The possibility of the breaking of the conformal invariance may provide a physical mechanism  allowing
to understand the small-scale structure of the gravitational gravity, and of the physics of the Planck scale. Based on this idea, a theory of gravity constructed from the assumption that  conformal symmetry is an exact local, but spontaneously broken
symmetry, was considered in \cite{Ho7}.

In the present paper we have considered another formulation of Weyl's theory, in which the scalar degree of freedom naturally appears within the framework of the theory, and is geometric in its origin. The introduction of the auxiliary geometric scalar field significantly simplifies the mathematical formalism, and leads to the linearization of the originally quadratic action in  the curvature scalar. This theory has been used to investigate various cosmological and astrophysical aspects, and in this investigation we have considered in detail the possibility that Weyl geometry could account for the observed dynamics of the galactic rotation curves.

Tentatively, to test this hypothesis, we have adopted an exact spherically symmetric solution of the Weyl geometric gravity theory as describing the geometry of the galactic halo, outside the baryonic matter distribution. We are of course aware of the limitations and simplification involved in this choice, since many other Weyl geometrical solutions may exist. However, as a first step in the direction of the Weylian description of "dark matter", this approximation may provide some hints for the viability/nonviability of such an approach. By using this exact solution we have obtained the full general relativistic expression of the tangential velocities of the massive particles in circular stable orbits, and we have used this expression, without ante-Newtonian or post-Newtonian approximations, to compare the theoretical model with the observational data. We have compared the Weyl geometric tangential velocity expression with the rotation curves of 171 galaxies of the SPARC sample. As one can see from Fig.~\ref{fig_chi2}, for 86 galaxies (50\% of the sample), the objective function $\chi^2$  has values smaller than one. For a number of 123 galaxies, the objective function took values smaller than 2. There are 139 galaxies (81\% of the sample) having $\chi^2<3$. Based on these statistical results, we may consider that the simplest possible Weyl geometric approach provides an acceptable description of the galactic rotation curves, indicating that more realistic approaches, based, for example, on the numerical solutions of the field equations, could improve the concordance between the model and the observational data.

We would also like to point out that by adding the baryonic matter contribution to the expression of the total velocity, the strict conformal symmetry of the model is broken. Adding matter in a conformally invariant way requires the consideration of the trace condition, as done, for example, in \cite{I5}. The inclusion of the matter would certainly modify the mathematical structure of the metric considered in the present study, and will open some new perspectives on the problem of the rotation curves.

One of the important requirements related to the motion of the particles in the galactic halo is that the timelike
circular geodesics must be stable. Let $r_{0}$ denote the radius of a
circular orbit. Let us now consider a small perturbation of the orbit $r_0$ of the form
$r=r_{0}+\delta $, where $\delta
<<r_{0}$ \citep{Lake, HaCh}. By expanding  $V_{eff}\left( r\right) $, as given by Eq.~(\ref{Veff})  and $%
\exp \left( \nu +\lambda \right) $ about $r=r_{0}$, it follows
from Eq.~(\ref{energy}) that the orbit perturbation satisfies the second order differential equation \cite{HaCh}
\begin{equation}
\ddot{\delta}(r)+\frac{1}{2}e^{\nu \left( r_{0}\right) +\lambda
\left( r_{0}\right) }V_{eff}^{\prime \prime }\left( r_{0}\right)
\delta (r)=0.
\end{equation}

Since for the present Weyl geometric gravity model the metric satisfies the condition $\nu +\lambda =0$, the condition for the stability of the  circular orbits requires that the condition $V_{eff}^{\prime \prime }\left( r_{0}\right) >0$ must be satisfied \cite{Lake, HaCh}. By assuming that $v_{tg}^2/c^2<<1$, from Eq.~(\ref{pot}) it follows that $V_{eff}(r)\approx e^{\nu (r)}$. By taking into account that $V'_{eff}(r)=\nu ' (r)e^{\nu (r)}=\left(2v_{tg}^2/c^2\right)e^{\nu}/r$, we obtain the stability condition of the circular orbits as
\be
\left|\frac{d}{dr} v_{tg}^2(r)-\frac{v_{tg}^2(r)}{r}+\frac{2v_{tg}^4}{c^2}\frac{1}{r}\right|_{r=r_0}>0.
\ee

By neglecting the last term in the above equation, the stability condition of the circular orbits can be formulated as
\be\label{cond1}
\left.\frac{d}{dr} v_{tg}^2(r)\right|_{r=r_0}>\frac{v_{tg}^2\left(r_0\right)}{r_0}.
\ee
If the inequality (\ref{cond1}) is true for $r_0 \in [R_b, R_{eff}]$
after dividing by $v^2(r_0)$ and integrating on this interval one obtains
the condition:
\be
v^2(R_{eff})/v^2(R_b)>R_{eff}/R_b.
\ee

As pointed out recently in \cite{Hob}, metrics of the type considered in the present investigation face the problem of their limiting tangential velocity, which, in the large radial coordinate limit tend to the speed of light, or other unrealistically high values. The impossibility of the existence of a plateau phase of the rotation curves has also been mentioned in this study. However, this raises the problem of the relation between mathematical and physical infinity. Indeed, in the limit $r\rightarrow \infty$ the tangential velocity (\ref{mfin}) tends to $c$. On the other hand, the range of the radial coordinate for realistic galactic halos is of the order of the few tenths of kpcs. In this range of $r$ a good description of the galactic rotation curves can be obtained. On the other hand, in the present model the radial dependence does appear via the dimensionless factor $r/C_2<<1$. The limit to the physical infinity $r\rightarrow \infty$ would require $r>>C_2 $, when already Weyl geometric effects are negligible. This can be easily seen from Eqs.~(\ref{40}) and (\ref{omeg}), which show that both the scalar field and the Weyl vector tend to zero, and we thus recover Einstein's general relativity. However, we may assume that the transition to general relativity occurs at some realistic finite astrophysical distances, $r=R$, so that for $r\geq R$, the metric is Schwarzschild, with $R$ obtained from the relation
\bea
e^{\nu (R) }&=&e^{-\lambda (R) }=1-\frac{3r_{g}}{C_{2}}-\frac{r_{g}}{R}+\left( 2-3%
\frac{r_{g}}{C_{2}}\right) \frac{R}{C_{2}}  \notag \\
&&+\left( 1+\frac{C_{1}}{12}-\frac{r_{g}}{C_{2}}\right) \frac{R^{2}}{%
C_{2}^{2}}\approx 1-\frac{R_g}{R},
\eea
where $R_g=GM_{tot}/c^2$, with $M_{tot}$ denoting the total galactic mass, including the contribution coming from the Weyl geometry. On the other hand we can introduce, by analogy with the Schwarzschild case, an effective dark matter gravitational radius $R_g^{(eff)}$, and an effective dark matter mass $M_{DM}^{(eff)}$, defined according to $e^{\nu (r)}=e^{-\lambda (r)}=1-R_g^{(eff)}(r)/r$, giving
\bea
R_g^{(eff)}&=&\frac{2GM_{DM}^{(eff)}(r)}{c^2}=r_g+\frac{3r_g}{C_2}r\nonumber\\
&&-\left(2-3\frac{3r_g}{C_2}\right)\frac{r^2}{C_2}+\left(1+\frac{C_1}{12}-\frac{r_g}{C_2}\right)\frac{r^3}{C_2^2},\nonumber\\
\eea
and an effective dark matter density $\rho _{DM}^{(eff)}(r)$, defined by using the standard expression  $\left(1/4\pi r^2\right)dM_{DM}^{(eff)}(r)/dr$, and which can be obtained as
\bea\label{56}
\rho _{DM}^{(eff)}(r)&=&\frac{c^2}{8\pi G}\Bigg[\frac{3r_g}{C_2}-2\left(2-\frac{3r_g}{C_2}\right)\frac{r}{C_2}\nonumber\\
&&-3\left(1+\frac{C_1}{12}-\frac{r_g}{C_2}\right)\frac{r^2}{C_2^2}\Bigg]\frac{1}{r^2}.
\eea

Since the ratio $r_g/C_2$ has negligibly small values, and $C_1$ has negative values, with $abs(C_1)>>1$, and $r/C_2<<1$, one can approximate Eq.~(\ref{56}) as
 \be
 \rho _{DM}^{(eff)}(r)\approx \frac{c^2}{8\pi G}\left[\frac{1}{4}\frac{C_1}{C_2^2}-\frac{4}{C_2r}\right].
 \ee
By taking now into account the expressions (\ref{40}) and (\ref{omeg}) of the scalar field and of the Weyl vector, after eliminating the constants we obtain the energy density of the effective Weyl geometric dark matter in the form
\be\label{58b}
 \rho _{DM}^{(eff)}(r)\approx \frac{c^2}{32\pi G}\left[\Phi (r)+8\alpha \frac{\omega _1(r)}{r}\right].
\ee

Eq.~(\ref{58b}) is valid in a region of space-time where the variation of the scalar field and of the Weyl vector is very slow. Hence, the effective physical properties of the galactic dark matter halos, including their density and mass distribution is indeed determined by the geometrical degrees of freedom that characterize the Weyl geometry effects at the galactic level.

The results of this investigation have provided some evidence for the potential of the simplest Weyl geometric gravity model as an alternative to the dark matter paradigm. Further, and detailed investigations in this field are certainly necessary to convincingly confirm, or infirm, the validity, and astrophysical relevance of this approach. Our results are thus only a first step in developing the theoretical and observational tools necessary to test the presence/absence of Weyl geometrical effects at the galactic and met-galactic levels.

\section*{Acknowledgments}

The work of TH is supported by a grant of the Romanian Ministry of Education
and Research, CNCS-UEFISCDI, project number PN-III-P4-ID-PCE-2020-2255
(PNCDI III).


\end{document}